\documentclass{aa}  

\usepackage{graphicx}
\usepackage{comment}
\usepackage{multirow}
\usepackage{booktabs}
\usepackage{tabularx}
\usepackage{xspace}
\usepackage{xcolor}
\usepackage[colorlinks]{hyperref}
\hypersetup{
    linkcolor=black,       % Color for internal links (e.g., table of contents, references)
    citecolor=black,      % Color for citations
    filecolor=magenta,    % Color for links to local files
    urlcolor=blue         % Color for URLs
}
\usepackage{txfonts}
\usepackage{siunitx}
\makeatletter
\let\linenumbers\relax
\let\@linenumberpar\relax
\makeatother
\begin{document} 

    \title{Secular brightness curves of 272 comets}
    
    \author{P. Lacerda \inst{1,7}
        \and A. Guilbert-Lepoutre\inst{2}
        \and R. Kokotanekova\inst{3}
        \and L. Inno\inst{4}
        \and E. Mazzotta Epifani\inst{5}
        \and C. Snodgrass\inst{6}
    }

    \offprints{\email{lacerda.pedro@gmail.com}}
    
    \institute{
        Instituto de Astrofísica e Ciências do Espaço, Universidade de Coimbra, Portugal\\ \email{lacerda.pedro@gmail.com}
        \and Laboratoire de G\'{e}ologie de Lyon: Terre, Plan\`{e}tes, Environnnement, CNRS, UCBL, ENSL, Villeurbanne, France
        \and Institute of Astronomy and National Astronomical Observatory, Bulgarian Academy of Sciences, 72 Tsarigradsko Chaussee Blvd., 1784 Sofia, Bulgaria
        \and Dep. Science and Technology, Parthenope University of Naples 
        \and INAF - Osservatorio Astronomico di Roma, Italy
        \and The University of Edinburgh, UK
        \and Instituto Pedro Nunes, Coimbra, Portugal
        \\
    }
    
    \date{Accepted March 24, 2025}

  \abstract
  {}
   {We investigate the brightening behavior of long-period comets as a function of dynamical age, defined by the original reciprocal semimajor axis, $1/a_0$. Our goal is to test long-standing claims about comet behavior using a large number of available measurements.}
   {We use a large set of photometric observations to compute and analyze global and local brightening curves for 272 long-period comets. Observed magnitudes are fitted with a linear model in log heliocentric distance, from which we derive brightening parameters for each comet. We categorize the sample into dynamically new, intermediate, and old comets, comparing their brightening behavior. We also examine the relationships between dynamical age and other orbital and physical parameters.}
   {Dynamically new comets are seen to brighten more slowly than old comets, particularly within 3 au from the Sun. The brightening rate of new comets appears to vary with heliocentric distance. New comets are intrinsically brighter than old comets, and exhibit a tighter correlation between brightening parameters.}
   {}


   \maketitle

    \section{Introduction}
    \label{section:introduction}

    The Oort Cloud is a vast, diffuse spherical shell composed of a trillion icy objects, extending from approximately a few thousand to tens of thousands of astronomical units (au) from the Sun \citep{1950BAN....11...91Oort, 2005ApJ...635.1348Francis}. The objects in the Oort Cloud are remnants from the early solar system that were scattered by planetary encounters and had their inclinations randomized and their perihelia raised far beyond the planetary region by a complex combination of Milky Way tides and perturbations by passing field stars \citep{1986Icar...65...13Heisler,1987Icar...70..269Heisler,1987AJ.....94.1330Duncan,2007AJ....134.1693Higuchi,2015AJ....150...26Higuchi,2024NatAs...8.1380Pfalzner}. The same processes cause some of these objects to return to the inner solar system, where they manifest as long-period comets \citep{1950BAN....11...91Oort,1986Icar...65...13Heisler,2009Sci...325.1234Kaib}.

    \citet{1950BAN....11...91Oort} identified an overabundance of long-period comets (LPCs) on orbits with extremely low reciprocal semimajor axis, specifically $1/a_0<0.0001$ au$^{-1}$, where $1/a_0$ corresponds to the energy of the comet's orbit. The semimajor axis, $a_0$, refers to the \textit{original} orbit before planetary perturbations. When comets pass through the planetary region, their orbits are perturbed, primarily by Jupiter, resulting in changes in reciprocal semimajor axis of order $\pm 0.0005$ au$^{-1}$ \citep{1948BAN....10..445VanWoerkom}. Oort took the observed abundance of $1/a_0$ values considerably smaller than this perturbation as evidence that these were dynamically new comets, coming into the planetary region for the first time. Such new comets are interesting because, unlike returning LPCs or short-period comets (SPCs), they have experienced minimal solar irradiation since their ejection from the planetary region.

    \begin{table*}
        \caption{Oort dynamical groups}
        \label{tab:oort-groups}
        \centering
        \begin{tabular}{@{}lccc@{}}
        \toprule
        Oort group &  $a_0$ (au)    &$1/a_0$ (au$^{-1}$)                  & $P$ (yr)         \\ \midrule
        new        &  $>$10,000     &$0$ to $10^{-4}$                       & $>10^6$          \\
        intermediate &  500 to 10,000 &$10^{-4}$ to $2\times 10^{-3}$         & 11,180 to $10^6$ \\
        old        &  25 to 500     &$2\times 10^{-3}$ to $4\times 10^{-2}$ & 125 to 11,180    \\
        periodic   &  $<$25         &$>4\times 10^{-2}$                     & $<$125           \\ \bottomrule
        \end{tabular}
        \tablefoot{Dynamical groups (column 1) defined in \citet{1951BAN....11..259Oort}, their original semi-major axis (column 2) and its reciprocal (column 3), and their range of orbital periods (column 4). Only the dynamically new, intermediate and old groups are used throughout the paper.}
    \end{table*}

    We build on the work of \citet{1951BAN....11..259Oort}, who first reported photometric differences between dynamically new and returning comets, namely that new comets brighten more slowly as they approach the Sun than returning comets.  Subsequent studies found evidence for the same behavior \citep{1976NASSP.393..410Meisel,1978M&P....18..343Whipple,1982come.coll..413Meisel,1995ICQ....17..168Green,1995Icar..118..223AHearn,2024PSJ.....5..273Holt}, but all used $\sim10$ times smaller samples or lacked coverage beyond 3 au. In this work, we employ a larger sample of LPCs observed mainly since the mid-1990s and to heliocentric distances beyond Jupiter, analyzing their brightening as they approach the Sun with the goal of testing the aforementioned claims about comet behavior. We examine specifically the \textit{secular} brightening behavior, setting aside the apparent unpredictability of comets, which often display large variations in behavior over short timescales and between individual objects.
    
    Even though previous studies vary in the details of how they measure comet brightness, they all attempt to track the comet \textit{total magnitude}, including light reflected by the nucleus and the (mostly dominant) contribution from the coma. Similarly, this study focuses on comet total magnitudes. As described in Section~\ref{section:methods}, although we use highly heterogeneous data, we follow a uniform strategy to compare brightening rates between comets. Distinct brightening behavior is indicative of differences in the mechanism of comet activity and linked to the physical properties of comet nuclei \citep{2004come.book..317Meech}.
    
    Oort's grouping of LPCs into \textit{new}, \textit{intermediate}, \textit{old}, and \textit{periodic} categories, based on the original orbit's reciprocal semi-major axis (see Table~\ref{tab:oort-groups}), remains a useful framework\footnote{We renamed Oort's \textit{fairly new} category as \textit{intermediate}.}. \textit{New} comets are those likely entering the inner solar system for the first time, \textit{intermediate} comets are those likely returning after a distant first perihelion, and \textit{old} comets have likely visited the planetary region multiple times. We may refer to \textit{newer} comets as encompassing new and intermediate comets or \textit{returning} comets when referring to intermediate and old comets together. \textit{Periodic} comets, which are primarily short-period comets (SPCs) are excluded from this study.

    SPCs, defined as having orbital periods less than 200 yr, include both Jupiter-family comets with low-inclination, prograde orbits strongly influenced by Jupiter, and Halley-type comets with nearly-isotropic orbits. Although spacecraft have visited several SPCs, revealing a diversity of physical characteristics and substantial processing of their nuclei \citep[e.g.,][and references therein]{2020SSRv..216...14Keller}, a similar mission to LPCs has yet to be achieved due to the logistical challenges posed by their long-period orbits. The upcoming Comet Interceptor (CI) mission, a joint endeavor by ESA and JAXA, aims to overcome this by pre-positioning a spacecraft at the Sun-Earth Lagrange point L2, ready to intercept a suitable LPC when discovered \citep{2024SSRv..220....9Jones}. The target comet for CI will likely be discovered by the Legacy Survey of Space and Time \citep[LSST;][]{2019ApJ...873..111Ivezic} to be carried out at the Vera Rubin Observatory, at distances well beyond Jupiter \citep{2024SSRv..220....9Jones}. The mission's preparation involves studying the detectability of potential targets and predicting their activity at flyby distances based on initial observations.

    \section{Data and methods}
    \label{section:methods}

    This Section describes the formalism, data sources, selection criteria, and analytical methods used to investigate the secular brightening behavior of LPCs. Our analysis begins with a comprehensive dataset of the over 3,000 known LPCs, which is then systematically refined through a series of selection steps based on specific criteria to ensure a robust final sample of comets for analysis. The selection process includes ensuring precise orbital solutions, exclusion of unbound comets and fragments, availability of photometric observations, application of quality criteria to address data uncertainty and non-uniformity. The details of these steps are provided in the following subsections. This approach allows us to focus the analysis on a well-defined set of 272 comets, examining their brightening behavior as a function of dynamical age and other orbital parameters.

    \subsection{Brightening curves}

    Comets brighten as they approach the Sun and the Earth. The brightening is due to a combination of observing geometry, as the comet approaches both the Sun and the Earth, and increasing back-scattering cross-section. We see comets in reflected or scattered sunlight, so the apparent brightness increases with the inverse square of both the heliocentric and geocentric distances. However, as comets approach the Sun, sublimation of their nuclei releases dust into an extended coma, increasing the reflective cross-section and leading to a steeper than inverse-square dependence on heliocentric distance. Furthermore, since the coma is not a point source, its apparent area increases as the geocentric distance ($\Delta$) decreases, leading to a shallower than inverse-square dependence on geocentric-distance \citep[the ``Delta effect'', e.g.][]{1993MNRAS.263..247Hughes}. Because this effect is small compared to other uncertainties, and in the interest of uniformity of treatment, we ignore it here.
    
    Expressed in magnitudes, the total magnitude ($T$) of a comet may be approximated by
    \begin{equation}
        T = \texttt{M1} + 5\log{\Delta} + \texttt{K1} \log{r} \,,
        \label{eq:total_magnitude}
    \end{equation}
    \noindent where $r$ and $\Delta$ are the heliocentric and geocentric distances in au, \texttt{M1} is the total magnitude at $r=\Delta=1$ au, \texttt{K1} is the brightening slope in units of mag per log au, and the factor $5$ assumes the total brightness depends solely on $\Delta^{-2}$. An inverse square dependence on heliocentric distance would imply $\texttt{K1}=5$, but increasing activity for incoming comets results in brightening slopes $\texttt{K1}>5$. 
    
    Parameters \texttt{M1} and \texttt{K1} are available from the NASA Jet Propulsion Laboratory (JPL) Solar System Dynamics\footnote{\href{https://ssd.jpl.nasa.gov}{\texttt{https://ssd.jpl.nasa.gov}}} (SSD) services for nearly all of the known comets. For LPCs, \texttt{K1} varies roughly between 4 and 48, with a median near 10 mag/log(au). The total magnitude parameter, \texttt{M1}, ranges from 4 to 22, with a median value of 10 magnitudes. \texttt{M1} is akin to the absolute magnitude, $H$, but applies to the total magnitude of active comets. Its range of validity is important when comparing to other studies.

    To remove the geocentric dependence, which oscillates on a shorter timescale than the cometary orbit, we also consider the heliocentric magnitude, $T_\odot = T-5\log(\Delta)$, given by
    \begin{equation}
        T_\odot = m + k \log{r} \,,
        \label{eq:heliocentric_magnitude}
    \end{equation}
    \noindent where $m$ and $k$ are parameters analogous to \texttt{M1} and \texttt{K1} in Eq.~\eqref{eq:total_magnitude}. In this paper, we use Eq.~\eqref{eq:heliocentric_magnitude} as the baseline for comparing brightening curves, focusing primarily on the parameter $k$, which we will refer to as \textit{brightening slope}. Parameter $m$ is the total heliocentric magnitude at $r=1$. The right-hand side of Equation~\ref{eq:heliocentric_magnitude} is also commonly written as $m+2.5n\log{r}$, where $n$ is referred to as ``activity index'' \citep{1988MNRAS.234..173Hughes,1992acm..proc..633Whipple}, or, less commonly, ``photometric index'' \citep{2011MNRAS.416..767Sosa}, or ``index of variation'' \citep{1976NASSP.393..410Meisel}. Parameter $k$ can be converted to the activity index as $n=k/2.5$.

    \subsection{Orbit quality selection}
    \label{subsec:data_methods_orbit_quality}

    Using the \texttt{astropy} package \texttt{astroquery} \citep{2019AJ....157...98Ginsburg}, we queried the Minor Planet Center (MPC) database to identify comet orbits, specifically selecting LPCs (prefix C/) that include original reciprocal semi-major axis values, $1/a_0$, and their uncertainties, $\sigma_{1/a_0}$. The MPC's $1/a_0$ values generally agree with those from other sources\footnote{As an example, of the 516 LPCs in the Nakano Note only 26 are classified differently using MPC orbits, 19 moving from intermediate to new, 5 moving the other way around, and 2 moving from intermediate to old. These differences have no impact on the conclusions presented in this paper.}, such as the CODE Catalog \citep{2020A&A...640A..97Krolikowska} and the Nakano Note website\footnote{\href{https://www.oaa.gr.jp/~oaacs/nk.htm}{\texttt{https://www.oaa.gr.jp/$\sim$oaacs/nk.htm}}} by Syuichi Nakano. Despite using more robust orbit calculations \citep{2024PSJ.....5..273Holt} these catalogs include only a subset of the orbital solutions available from the MPC. To maximize the sample size and ensure internal consistency, we adopted the MPC original orbits. The $1/a_0$ values were used to classify LPCs into the different Oort groups (see Table~\ref{tab:oort-groups}). We kept new, intermediate and old comets, and excluded periodic and unbound orbits. On the latter, we retained only comets with $1/a_0 > 3\sigma_{1/a_0}$, ensuring a $3\sigma$ confidence level that the orbits are bound ($1/a_0 > 0$). This criterion minimizes the likelihood of including unbound interlopers. Comet fragments were also excluded resulting in a preliminary sample of 787 comets for further inspection.
    
    We subsequently used the JPL SSD Small-Body Database (SBDB) Query API to retrieve additional parameters and associated uncertainties, $\sigma$, for each comet, including semi-major axis ($a, \sigma_a$), time of perihelion passage ($t_p, \sigma_{t_p}$), perihelion distance ($q, \sigma_q$), eccentricity ($e, \sigma_e$), inclination ($i, \sigma_i$), and the brightening and magnitude parameters (\texttt{M1} and \texttt{K1}). The latter result from fitting Eq. \eqref{eq:total_magnitude} to all MPC-reported observations. Importantly, in the case of \texttt{M1} and \texttt{K1}, this is done without distinguishing between incoming and outgoing data (see Section~\ref{subsec:jplvspaper} for more details).

    \begin{figure}
    \resizebox{\hsize}{!}{\includegraphics{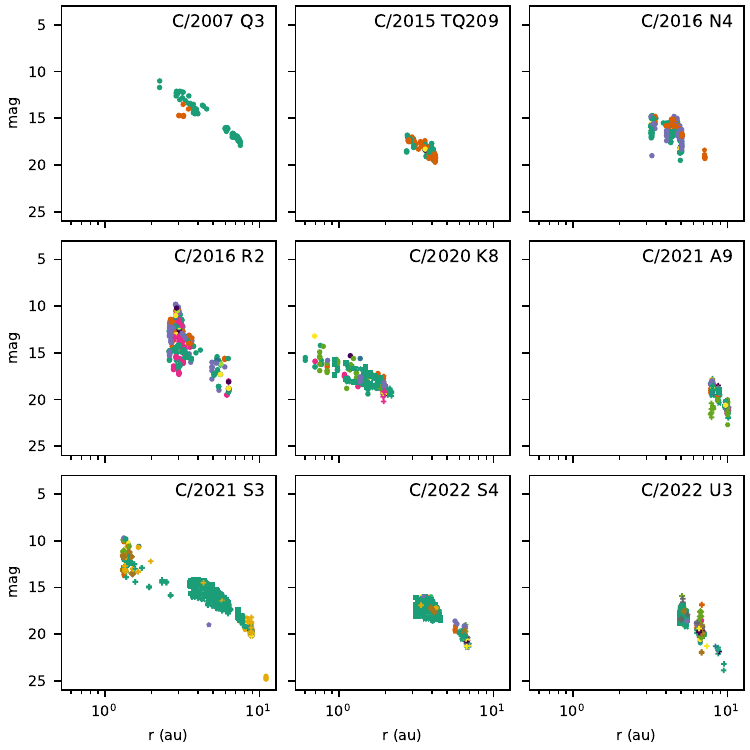}}
            \caption{Small subset of observations grouped by comet. Colors correspond to different observatories and symbols to different photometric bands. Larger sample in Appendix \ref{app:curves}.}
            \label{fig:obs_sample}
    \end{figure}

    \begin{table}
        \caption{Date and geometry of observations analyzed}
        \label{tab:observations}
        \centering
        \begin{tabular}{llll}
        \toprule
             & Date    & $r$   & $\Delta$ \\ \midrule
        min  & 1886.09 &  0.31 &  0.14     \\
        25\% & 2015.69 &  2.60 &  2.15     \\
        50\% & 2020.60 &  4.03 &  3.63     \\
        75\% & 2022.56 &  5.74 &  5.38     \\
        max  & 2024.60 & 34.62 & 34.33   \\ \bottomrule
        \end{tabular}
        \tablefoot{Minimum, maximum, median (50\%) and quartiles (25\% and 75\%) are listed for the date and geometry ($r$ and $\Delta$ are heliocentric and geocentric distance) distributions of the 247,773 observations analyzed.}
        \vspace{1pt}
    \end{table}

    \begin{table*}
        \caption{Number of strands, global curves and comets analyzed}
        \label{tab:strand_global_count}
        \centering
        \begin{tabular}{cccccccccccc}
        \toprule
         \multicolumn{6}{c}{992 strands (194 comets)} & \multicolumn{6}{c}{346 global curves (254 comets)} \\ \cmidrule(lr){1-6} \cmidrule(lr){7-12}
         \multicolumn{3}{c}{644 pre-perihelion (141 c.)} & \multicolumn{3}{c}{348 post-perihelion (112 c.)} & \multicolumn{3}{c}{198 pre-perihelion (198 c.)} & \multicolumn{3}{c}{148 post-perihelion (148 c.)} \\ \cmidrule(lr){1-3} \cmidrule(lr){4-6} \cmidrule(lr){7-9} \cmidrule(lr){10-12}
         new & int. & old & new & int. & old & new & int. & old & new & int. & old \\
         321 (61) & 238 (56) & 85 (24) & 154 (46) & 136 (42) & 58 (24) & 85 (85) & 71 (71) & 42 (42) & 54 (54) & 58 (58) & 36 (36) \\ 
        \bottomrule
        \end{tabular}
        \tablefoot{Numbers of strands and global curves split into orbital arc (pre- or post-perihelion) and Oort group. Number of comets included in each subgroup is shown in brackets. Each comet may have multiple strands but at most two global curves (1 pre- and 1 post-perihelion).} 
        \vspace{1pt}
    \end{table*}

    \begin{table}
        \caption{Number of comets per orbital arc, Oort group and strand count}
        \label{tab:comet_distribution_per_arc_oort_strands}
        \centering
        \begin{tabular}{llrrr}
        \toprule
                         & & \multicolumn{3}{c}{number of strands} \\ \cmidrule(lr){3-5}
         orbital arc     & Oort group   & 1  & 2-10 & 11+ \\ \midrule
         pre-perihelion  & new          & 22 & 34   & 5 \\
                         & intermediate & 19 & 32   & 5 \\
                         & old          &  5 & 18   & 1 \\
         post-perihelion & new          & 19 & 24   & 3 \\
                         & intermediate & 23 & 16   & 3 \\
                         & old          & 11 & 13   & 0 \\
        \bottomrule
        \end{tabular}
        \tablefoot{Each entry tallies how many comets have pre- or post-perihelion data, of which Oort group and with how many strands. E.g., there is only 1 dynamically old comet with more than 10 strands pre-perihelion.}
        \vspace{1pt}
    \end{table}

    \begin{table}
        \caption{Heliocentric coverage (in au) of strands}
        \label{tab:strand_geometry}
        \centering
        \begin{tabular}{lrrrrr}
        \toprule
         & \multicolumn{2}{r}{pre-perihelion strands} & \multicolumn{2}{r}{post-perihelion strands} \\ \cmidrule(){2-3} \cmidrule(lr){4-5}
         & midpoint & $r$ range & midpoint & $r$ range \\ \midrule
        min.    &  0.56 &  0.50 &  0.72 &  0.50 \\
        25\%    &  2.68 &  1.39 &  2.57 &  1.17 \\
        50\%    &  3.90 &  2.35 &  4.16 &  2.04 \\
        75\%    &  5.37 &  3.76 &  5.33 &  3.11 \\
        max.    & 15.05 & 19.87 & 10.09 & 10.21 \\
        \bottomrule
        \end{tabular}
        \tablefoot{Statistics of heliocentric distance and range (in au) of the strands analyzed. Strand midpoints were calculated in log space since fitting is linear in $\log{r}$.}
        \vspace{1pt}
    \end{table}

    \subsection{Magnitude data selection}
    \label{subsec:data_methods_data_selection}

    To analyze the total magnitude evolution of the comets in the preliminary sample identified in Subsection \ref{subsec:data_methods_orbit_quality} as they approach the Sun, we obtained magnitude data from the MPC. Using \texttt{astroquery}, we obtained all reported observations for each comet from the MPC observations database\footnote{\href{https://minorplanetcenter.net/db\_search}{\texttt{https://minorplanetcenter.net/db\_search}}}. These include magnitude, but are primarily used to measure the position and motion of the comets. Historically, the magnitude was specified simply as either nuclear (N) or total (T), but recent T magnitude data are often filter-specific. We focused on T magnitudes, considering observations taken through any available filter, but excluding N data. We chose to exclude N magnitudes for two primary reasons. Firstly, our focus is on the brightening behavior due to cometary activity, which can only be reliably accounted for in T magnitudes. Secondly, and more crucially, there is no straightforward way to verify whether the reported N magnitudes have been processed by observers to remove residual coma effects. Such processing, if present, could introduce additional systematic biases into our analysis. By focusing solely on total magnitudes, we aim to maintain consistency and minimize potential sources of systematic error. Comets for which \texttt{astroquery} returned no magnitudes, or only N magnitudes, were filtered out at this stage.

    Each MPC observation includes the comet's designation, the observation date and time, the observed magnitude, the photometry type (T or filter band pass), and the observatory code. We note that MPC magnitude uncertainties are not provided. To address the lack of uncertainty information, we developed a strategy which relies on sequences of consecutive measurements taken by the same observer instead of individual measurements. Each sequence allows determination of slope and respective uncertainty. This strategy is described in Subsection~\ref{subsec:data_methods_strands_as_tool}.
    
    We used the JPL Horizons File API to retrieve the observing geometry, specifically the heliocentric ($r$) and geocentric ($\Delta$) distances for each observation.  Additionally, we classified each observation as pre-perihelion or post-perihelion based on the observation time relative to the perihelion passage.

    The final dataset consists of 247,773 measurements for 741 comets, obtained at 978 different observatories, spanning heliocentric distances from 0.3 to 35 au. Most measurements are from survey observatories, including ATLAS \citep{2020PASP..132h5002Smith}, Pan-STARRS \citep{2004AN....325..636Hodapp}, the Catalina Sky Survey \citep{2003DPS....35.3604Larson}, and LINEAR \citep{2000Icar..148...21Stokes}. The majority of the data (over $99\%$) were collected since 1996, thanks to these surveys and contributions from amateur observers at various observatories across Europe, notably at Tarbatness Observatory (Portmahomack, Scotland), Olmen (Balen, Belgium), Obs. Chante-Perdrix (Dauban, France), Brixiis Observatories (Kruibeke, Belgium), and Grömme (Oudsbergen, Belgium). A sample of observations grouped by comet is shown in Figure~\ref{fig:obs_sample}.  Table~\ref{tab:observations} describes the dates and distances of the measurements. The full table of observations is made available online \href{https://doi.org/10.5281/zenodo.15100031}{here}.

    \subsection{Robust brightening measurement using strands}
    \label{subsec:data_methods_strands_as_tool}

\begin{figure*}
        \centering
        \includegraphics[width=17cm]{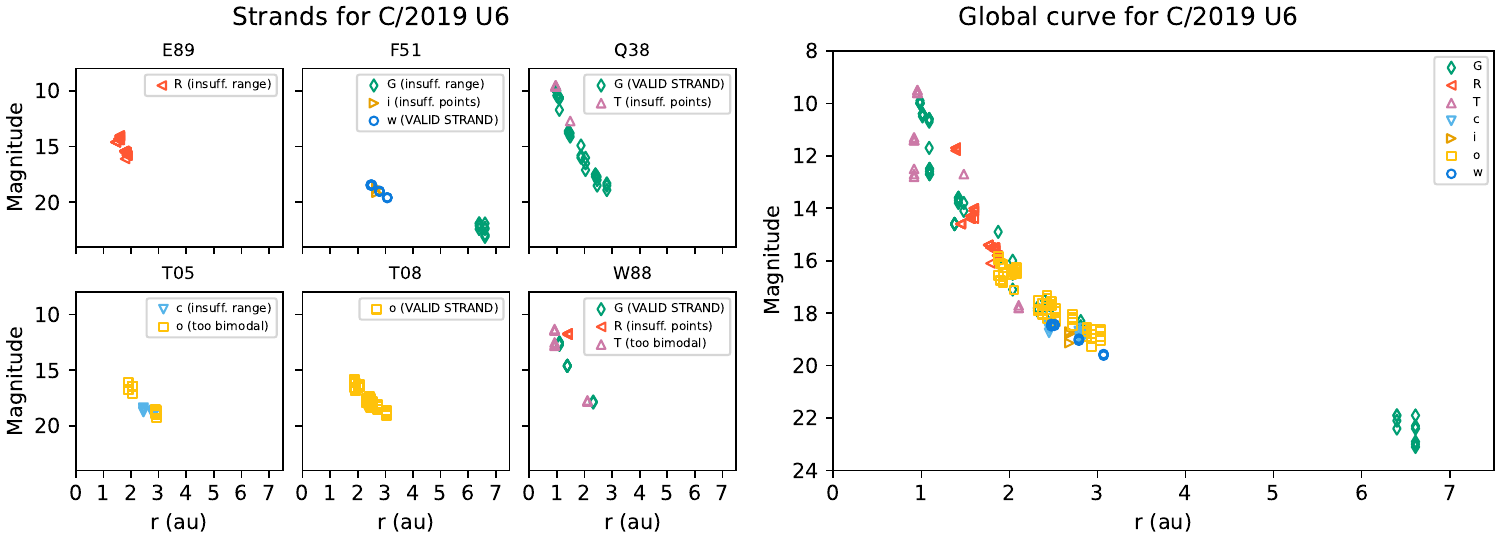}
        \caption{Strands (top panel array) and global curve (bottom panel) for comet \object{C/2019 U6}. Top array panels are labeled by MPC observatory code (E89=Geyserland, F51=Pan-STARRS 1, Q38=Lake Boga, T05=ATLAS-HKO, T08=ATLS-MLO, W88=Slooh). Top-panel legends indicate strand filter and whether strands are valid or invalid, the latter indicated as the reason for rejection. Global curve legend indicates filter (same symbols as top array panels), for comparison. Note that the global curve may include points from invalid strands, e.g., green diamonds in top sub-panel F51, which means they were not useful to assess local slope, but are useful to assess the global behavior.}
        \label{fig:strands_and_global_curve}
    \end{figure*}

    The MPC magnitude data described above were collected mainly for astrometry and orbit determination and hence the lack of photometric uncertainties. Furthermore, comet photometry is inherently complex \citep{1991ASSL..167...19Jewitt} and is often influenced by subjective factors, which are not detailed for the MPC observations used in this study. Variations in observational techniques (e.g., aperture selection, different filters) may introduce magnitude offsets between observatories. As a result, the data for each comet are highly heterogeneous, presenting significant challenges for analysis. To address these challenges, we introduce \textit{strands}: sequences of consecutive measurements taken by the same observatory using the same filter. It may be helpful to inspect Figure~\ref{fig:strands_and_global_curve} for a visual representation of strands.

    We fitted Eq.~\eqref{eq:heliocentric_magnitude} to separate strands, to extract their slope and magnitude parameters, which we will denote $k_r$ and $m_r$. A key advantage of this approach is that strand-based slopes ($k_r$) are robust to differences between observers and to color offsets because each strand is derived from a single filter at a given observatory, and their slope depends only on relative changes within a single strand. This ensures that the brightening trends we extract are as free as possible from systematic observational biases. Each strand also provides an uncertainty estimate on $k_r$ and $m_r$, denoted  $\sigma_{k_r}$ and $\sigma_{m_r}$, allowing us to assess the quality of the data and filter out unreliable strands (discussed in the next Subsection). We note that $m_r$ is still susceptible to variations in observational techniques, such as aperture selection or calibration inconsistencies, so its interpretation requires more caution.
    
    In the subsequent analysis, strands serve as the fundamental building blocks for studying the comet’s brightening behavior. Another important benefit of strands, is that they represent different heliocentric distances; this is highlighted by the subscript $r$ in $k_r$ and $m_r$. As will become clear below, this enables us to measure changes in brightening rate with heliocentric distance.
    
    \begin{figure}
        \resizebox{\hsize}{!}{\includegraphics{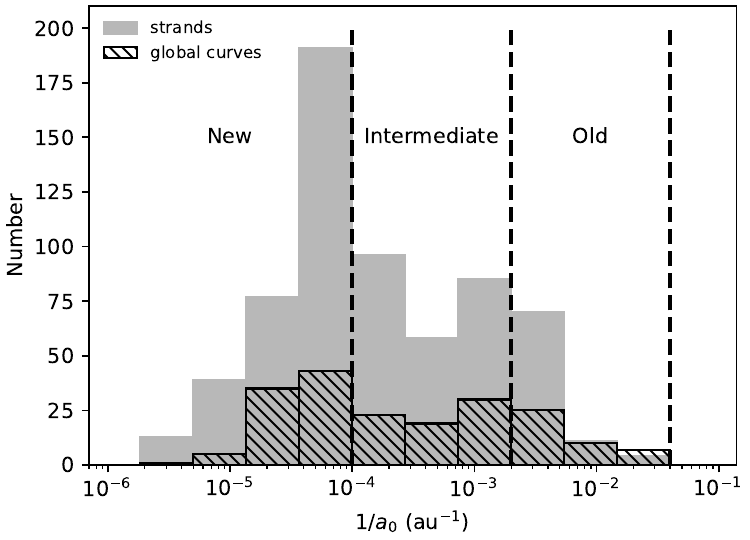}}
        \caption{Distribution of $1/a_0$ for strands (multiple per comet) and global curves (one or two per comet). Oort group ranges are identified.}
        \label{fig:hist_recipa0}
    \end{figure}

    \begin{figure*}
        \centering
        \includegraphics[width=17cm]{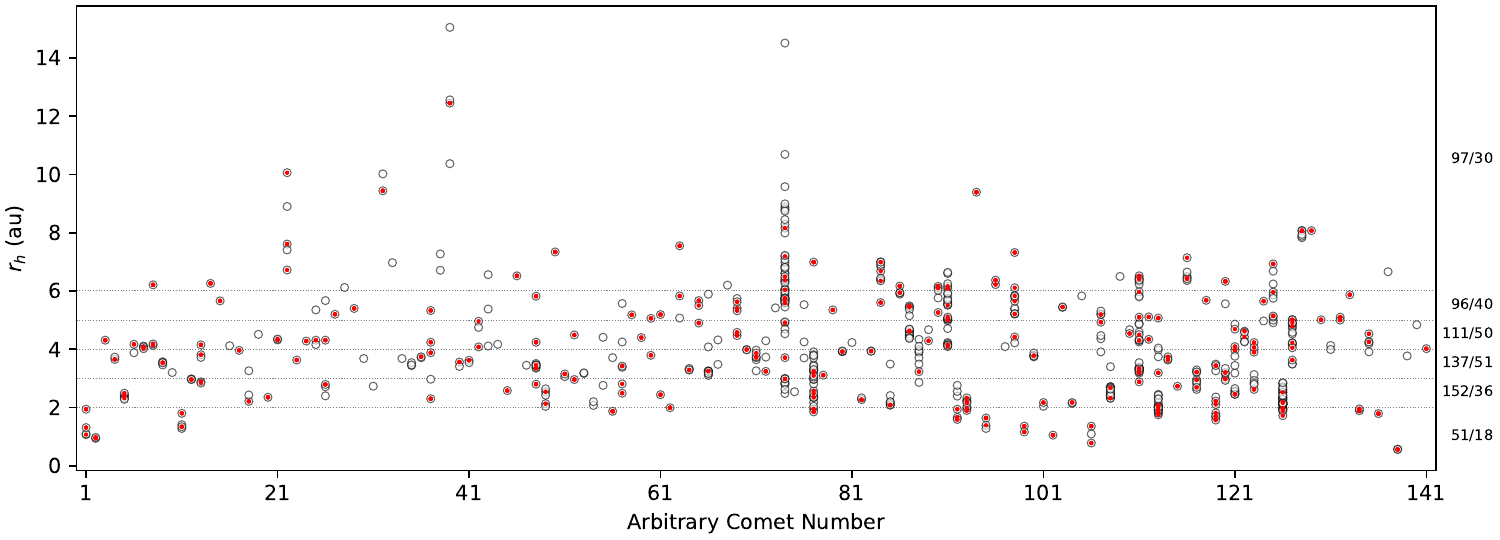}
        \caption{Pre-perihelion strand midpoints (vertical axis) are plotted as open circles for each comet (horizontal axis) in the original sample (644 strands for 141 comets). Horizontal dotted lines mark heliocentric bin boundaries. Numbers of strands/comets in each bin in the original sample are shown to the right of the Figure. A debiased strand sample (see text for details) is overplotted as red dots.}
        \label{fig:strands_per_comet_rh}
    \end{figure*}

    \subsection{Strand quality selection}
    \label{subsec:data_methods_strand_quality}

    To ensure that strand-based brightening rate measurements are reliable, we first applied a set of selection criteria before performing any slope fitting. The fundamental requirement is that a strand must provide a reliable estimate of the brightening slope, $k_r$. If a strand lacks sufficient data points, does not span a sufficient range in heliocentric distance, or consists primarily of clustered measurements at its extremities, its slope may be misleading. Thus, to be considered reliable, a strand must satisfy the following conditions:
    \begin{itemize}
        \item It should contain at least 10 individual measurements;
        \item It should span at least 0.5 au in heliocentric distance;
        \item Its measurements should be well-distributed in heliocentric distance, rather than forming isolated clusters at the extremities. This was enforced by binning the data into five equal intervals in  $\log{r}$  and requiring at least one measurement outside bins 1 and 5.
    \end{itemize}
    These particular criteria are based on extensive visual and numerical inspection of the resulting fits. Strands with fewer measurements or insufficiently broad or bimodal coverage often yielded fits that, while sometimes statistically acceptable, did not reliably reflect the comet's secular brightening trend. More complex, adaptable criteria, produced equivalent results, so we opted to use a simpler and more easily reproducible set of thresholds.
    
    Our focus is on the secular behavior of comets, so we evaluated strands based on the expectation of regular brightening, excluding data that did not conform to this pattern. By ``regular brightening,'' we mean that a comet's brightness should vary according to Eq.~\eqref{eq:heliocentric_magnitude}. Spurious data would introduce scatter that disrupts this smooth trend. Thus, to assess each comet's secular brightening, we aimed to reject data that exhibit excessive scatter. Although comet outbursts can complicate this assessment, filtering out such events is acceptable, as this study focuses on long-term brightening trends.
    
    As such, we subjected strands that meet the criteria above to an ordinary least squares (OLS) fit to Eq.~\eqref{eq:heliocentric_magnitude}, and used the F-test $p$-value to reject the null hypothesis that all fit parameters are zero, i.e. that the strand magnitudes are consistent with noise. The F-test is well-established and easy interpret when assessing linearity: a large $p$-value indicates that Eq.~\eqref{eq:heliocentric_magnitude} has no explanatory power regarding the strand magnitudes. Only strands with $p_\text{val}<0.0027$ ($3\sigma$) and $k_r$ uncertainties $\sigma_{k_r} < 1$ were considered valid.

    \subsection{Extraction of strand brightening parameters}
    \label{subsec:data_methods_extraction_brightening_parameters}

    We extracted the brightening parameters $m_r$ and $k_r$ and respective uncertainties by fitting Eq.~\eqref{eq:heliocentric_magnitude} to each of the valid strands using a robust linear model (RLM\footnote{RLM works by iteratively weighting down outliers following a specific function. We use Python's \texttt{statsmodels} implementation of RLM with the default outlier weighting function \texttt{HuberT} \citep{1981Robuststatistics.Huber}.}). We chose to use RLM to extract the brightening parameters for analysis because it is more effective in capturing the main slope in the presence of outliers. We entertained the possibility that differences in number and density of measurements per strand could affect the robustness of their respective $k_r$ values, but tests involving the most extreme cases showed no significant deviation from the initial RLM fits.
    
    Finally, each strand was associated with its comet, observatory, and photometric band, along with the comet's original orbital properties, perihelion, and discovery information. Additionally, the strand's specific properties, including number of measurements, minimum and maximum heliocentric distances ($r$), midpoint in $\log{r}$, and best-fit parameters $k_r$ and $m_r$, were recorded.

    \subsection{Global curves}
    \label{subsec:data_methods_global_curves}
    
    To further understand the overall brightening trends of each comet, we performed two additional Eq.~\ref{eq:heliocentric_magnitude} fits per comet: one using all pre-perihelion data and the other using all post-perihelion data, irrespective of the observatory or photometric band. These \textit{global fits} and associated best-fit parameters, which we will denote $k_1$ and $m_1$, allowed us to measure the overall slope of each comet's brightening curve. Valid global curves were also required a minimum of 10 measurements covering at least 0.5 au in a relatively uniform manner, the same fit quality criteria as individual strands, and the inclusion of at least two strands.

    \subsection{Sample summary and debiasing}
    \label{subsec:data_methods_sample_summary}
    
    The resulting sample includes 272 comets, 176 with both valid strands and global curves, 78 with global curves but no valid strands and 18 with valid strands but no global curves. In total, 992 strands were deemed fit for analysis, corresponding to 194 comets. Approximately one-third of the comets have only a single strand, with a median of two strands per comet, and a maximum of 67 strands for comet \object{C/2017 K2}. Most of the strands are pre-perihelion (644) compared to post-perihelion (348). As for global curves, 254 comets have valid coverage of either pre-perihelion brightening or post-perihelion fading, while 92 comets have both pre- and post-perihelion global curves. Tables \ref{tab:strand_global_count} and \ref{tab:comet_distribution_per_arc_oort_strands} summarize how comets, strands and global curves distribute over orbital arc and Oort group. Figure~\ref{fig:hist_recipa0} shows the distribution of $1/a_0$ for global curves and strands. Figure~\ref{fig:strands_per_comet_rh} and Table~\ref{tab:strand_geometry} show how strands are distributed in heliocentric distance. The resulting table of strands and global curves and their properties is made available online \href{https://doi.org/10.5281/zenodo.15100031}{here}.

    In the specific case of strands, the sample summarized above is biased because some comets contribute more strands than others. We could address this problem by randomly resampling the strand sample ensuring each comet contributes the same amount of strands. However, this would lead to uneven coverage of heliocentric distance. As we will see below, the brightening slope, $k_r$, appears to vary with heliocentric distance, making it important to ensure that all distance ranges are evenly represented. To address the biases, we first binned strands according to their midpoint in log heliocentric distance, which serves as an anchor for each strand when modeled using Eq.~\ref{eq:heliocentric_magnitude}. The bin boundaries were set at 2, 3, 4, 5 and 6, extending indefinitely at both ends, attempting uniform coverage of the range containing the most data. We experimented with different bin structures, but found no significant impact in the results as long as each bin contains at least 30 strands. Next, we randomly sampled (with replacement) an equal number of strands from each bin, while simultaneously ensuring each comet contributes the same amount of strands. As an example for the case of pre-perihelion strands, we selected 107 or 108 strands per bin, ensuring that the total number of resampled strands remains 644. In the bin nearest the Sun, we drew 5 or 6 strands for each of the 18 comets, obtaining a total of 107 resampled strands for the bin (see Figure~\ref{fig:strands_per_comet_rh}). The resulting samples, which we will refer to as \textit{debiased strand samples}, are a fairer representation of the variability due to individual comets and heliocentric distances.
    
    \begin{figure}
        \resizebox{\hsize}{!}{\includegraphics{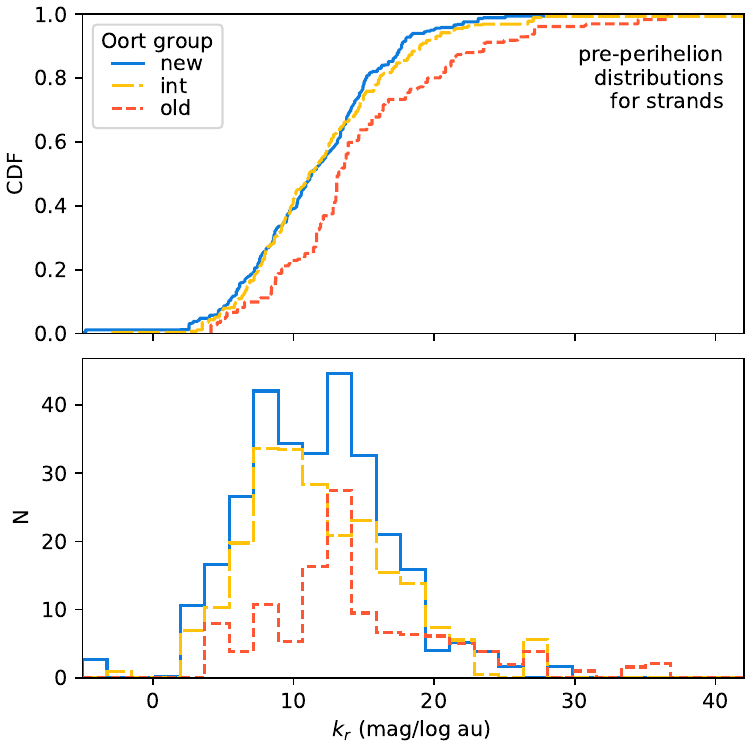}}
        \caption{Cumulative probability distributions (top) and histograms (bottom) of the $k_r$ brightening parameter for pre-perihelion strands, by Oort group. Shown are averages of $N_i=100$ debiased strand sample distributions (see text for details). Histogram counts were rescaled so that they add up to the actual number of pre-perihelion strands measured.}
        \label{fig:ecdf_kr_strands}
    \end{figure}

    \begin{figure}
        \resizebox{\hsize}{!}{\includegraphics{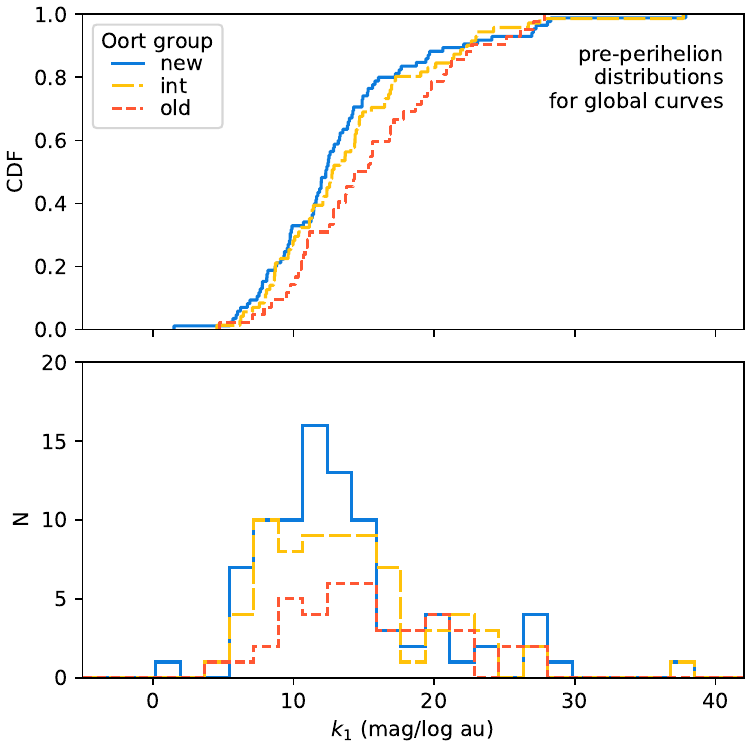}}
        \caption{Same as Figure~\ref{fig:ecdf_kr_strands} but for the $k_1$ brightening parameter fit to pre-perihelion global curves.}
        \label{fig:ecdf_k1_global}
    \end{figure}

    \begin{table*}
        \caption{Strand parameter statistics by Oort group and orbital arc}
        \label{tab:sample_stats_kr_mr_strands}
        \centering
        \begin{tabular}{lcccccc}
        \toprule
        \multicolumn{7}{l}{Pre-perihelion strands} \\ \midrule
                   & \multicolumn{3}{c}{$k_r$ statistics}       & \multicolumn{3}{c}{$m_r$ statistics}       \\ \cmidrule(lr){2-4} \cmidrule(lr){5-7}
                   & new           & int.          & old           & new           & int.          & old           \\ \cmidrule(lr){2-4} \cmidrule(lr){5-7}
        $N$        & 298.4         & 227.7         & 117.9         & 298.4         & 227.7         & 117.9         \\
        mean \& SE & $11.51\pm0.03$& $12.05\pm0.04$& $14.80\pm0.06$& $8.17\pm0.03$ & $8.79\pm0.03$ & $10.97\pm0.03$\\
        25\%       & 7.81          & 7.99          & 10.82         & 5.55          & 6.14          & 8.84          \\
        median     & 11.30         & 11.11         & 13.22         & 8.45          & 9.23          & 11.99         \\
        75\%       & 14.86         & 15.00         & 18.28         & 10.56         & 12.10         & 13.42         \\ \midrule
        \multicolumn{7}{l}{Post-perihelion strands} \\ \midrule
                   & \multicolumn{3}{c}{$k_r$ statistics}       & \multicolumn{3}{c}{$m_r$ statistics}       \\ \cmidrule(lr){2-4} \cmidrule(lr){5-7}
                   & new           & int.          & old           & new           & int.          & old           \\ \cmidrule(lr){2-4} \cmidrule(lr){5-7}
        $N$        & 156.6       & 124.4           & 67.0          & 156.6         & 124.4        & 67.0            \\
        mean \& SE & $12.55\pm0.04$& $11.94\pm0.04$& $13.83\pm0.06$& $7.94\pm0.02$ &$8.72\pm0.0.04$& $9.79\pm0.04$ \\
        25\%       & 9.61          & 9.51          & 10.28         &  5.76         &  6.84         &  7.03         \\
        median     & 12.05         & 11.59         & 13.63         &  7.89         &  9.34         & 10.32         \\
        75\%       & 14.38         & 13.82         & 16.56         & 10.04         & 11.41         & 12.04         \\ 
        \bottomrule
        \vspace{1pt}
        \end{tabular}
        \tablefoot{Statistics of $k_r$ and $m_r$ grouped by Oort group (new, intermediate and old comets) and orbital arc (pre- and post-perihelion). Statistical quantities listed are averages over 100 debiased strand samples (each may have different Oort group numbers). Mean, standard error on the mean (SE), median and 25\% and 75\% quartiles are in units of mag/$\log{\text{au}}$.}
    \end{table*}

    \begin{table}
        \caption{Bootstrap comparisons of strand and global curve parameters across Oort groups and orbital arcs}
        \label{tab:pairwise_prob_kr_mr_strands}
        \centering
        \begin{tabular}{lrrrr}
        \toprule
        Pre-perihelion& \multicolumn{2}{c}{Strands}                                                 & \multicolumn{2}{c}{Global curves}                                     \\ \cmidrule(lr){1-1}\cmidrule(lr){2-3}\cmidrule(lr){4-5}
                      & \multicolumn{1}{c}{$k_r$} & \multicolumn{1}{c}{$m_r$} & \multicolumn{1}{c}{$k_1$} & \multicolumn{1}{c}{$m_1$} \\ \cmidrule(lr){2-2}\cmidrule(lr){3-3}\cmidrule(lr){4-4}\cmidrule(lr){5-5}
        new vs.\ int. & $0.1539$                             &  $0.0137$                            & $0.4491$                          &  $0.5190$                         \\
        old vs.\ int. & $0.0014$                             &  $0.0007$                            & $0.2343$                          &  $0.0135$                         \\
        new vs.\ old  & $0.0019$                             & $<0.0005$                            & $0.0255$                          &  $0.0015$                         \\ \midrule
        Post-perihelion& \multicolumn{2}{c}{Strands}                                 & \multicolumn{2}{c}{Global curves}                     \\ \cmidrule(lr){1-1}\cmidrule(lr){2-3}\cmidrule(lr){4-5}
                      & \multicolumn{1}{c}{$k_r$} & \multicolumn{1}{c}{$m_r$} & \multicolumn{1}{c}{$k_1$} & \multicolumn{1}{c}{$m_1$} \\ \cmidrule(lr){2-2}\cmidrule(lr){3-3}\cmidrule(lr){4-4}\cmidrule(lr){5-5}
        new vs.\ int. & $0.2091$                             & $0.0271$                             & $0.3402$                          &  $0.2463$                         \\
        old vs.\ int. & $0.0412$                             & $0.0824$                             & $0.7957$                          &  $0.0085$                         \\
        new vs.\ old  & $0.0607$                             & $0.0015$                             & $0.1923$                          & $<0.0005$                         \\
        \bottomrule
        \vspace{1pt}
        \end{tabular}
        \tablefoot{Shown are the $p$-value probabilities for the null hypothesis that the strands and global curves samples for new, intermediate, and old comets are drawn from the same parent populations. Pairwise comparisons (rows) were performed using $D_\text{KS}$ on strand parameters $k_r$ and $m_r$, and global curve parameters $k_1$ and $m_1$ (columns). Strand $p$-values are averages of 100 debiased strand samples. Lower $p$-values indicate stronger evidence against the null hypothesis, suggesting differences between Oort groups. Pre- and post-perihelion data were considered separately.}
    \end{table}

    \section{Results}

    \subsection{Comparing Oort groups using bootstrap}
    \label{subsec:bootstrap_pvals}
    To assess whether the photometric behaviors of new, intermediate, and old comets differ, we compared their brightening parameters by testing the null hypothesis that they are drawn from the same parent population. Using the debiased strand samples described above, we compiled $k_r$ and $m_r$ distributions for each Oort group, both pre- and post-perihelion. Since global curves do not require debiasing, we used used the $k_1$ and $m_1$ samples directly.
    
    We then applied bootstrap resampling to evaluate how often the observed differences between two Oort groups could arise by chance. Specifically, we constructed synthetic samples by randomly shuffling Oort group labels and re-computing the difference multiple times. This allowed us to estimate the probability ($p$-value) that a difference as large as the observed one would occur if the groups were inherently indistinguishable, ensuring robustness to sample size. We adopt a $3\sigma$ significance threshold, considering results statistically significant if $p_\text{val}<0.0027$.
    
    To quantify the differences between samples, we used the Kolmogorov-Smirnov $D$ statistic\footnote{Implemented using Python's \texttt{scipy.stats.ks\_2samp} function \texttt{ks\_2samp}.} ($D_\text{KS}$), which measures the maximum difference between the respective cumulative distributions. $D_\text{KS}$ is equal to zero when the compared distributions are identical, and increases as they diverge in location, scale, or shape. Our bootstrap approach and choice of metric are non-parametric as we are dealing with asymmetric distributions, and is robust to sample size. Since debiased strand samples are drawn at random from the original sample, which introduces variability, we generated multiple ($N_i$) instances and averaged the results to obtain stable estimates of statistical properties and $p$-values. We found that $N_i=100$ was sufficient achieve convergence.

    \subsection{Comparing brightening using strands}
    \label{subsec:results_kr_strands}

    Figure~\ref{fig:ecdf_kr_strands} shows cumulative distributions and histograms of the $k_r$ parameter, grouped by Oort group, and Table~\ref{tab:sample_stats_kr_mr_strands} details the statistical properties of each distribution. We subjected each debiased $k_r$ sample to the bootstrap resampling described at the start of this Section to calculate the $p$-values for the null hypothesis that the samples being compared are drawn from the same parent population. These are shown in Table~\ref{tab:pairwise_prob_kr_mr_strands}, where each $p$-value is the average of $N_i=100$ debiased strand samples.

    The distributions of brightening slopes for strands of new and intermediate comets are not significantly different ($p_\text{val}\approx 0.15$), both groups having median $k_r$ near 11.1 to 11.3 mag/$\log(\text{au})$. In contrast, old comets display markedly different behavior, characterized by a higher median $k_r=13.2$ and a distribution skewed to larger $k_r$, lacking comets that brighten very slowly. The $k_r$ distribution for old comets differs significantly from both new comets ($p_\text{val}=0.0019$) and intermediate comets ($p_\text{val}=0.0014$). Although \citet{1951BAN....11..259Oort} use a different parametrization of brightening with heliocentric distance \citep[for details, see][]{1970AJ.....75..252Meisel}, our results corroborate their primary conclusion that new comets display different photometric behavior, brightening more slowly as they approach the Sun than old comets.

    \begin{table*}
        \caption{Statistics of global curves $k_1$ and $m_1$ samples}
        \label{tab:sample_stats_k1_m1_global_curves}
        \centering
        \begin{tabular}{lcccccc}
        \toprule
        \multicolumn{7}{l}{Pre-perihelion global curves}                                                          \\ \midrule
                      & \multicolumn{3}{c}{$k_1$ statistics}          & \multicolumn{3}{c}{$m_1$ statistics}      \\ \cmidrule(lr){2-4} \cmidrule(lr){5-7}
                      & new           & intermediate  & old           & new         & intermediate & old          \\ \cmidrule(lr){2-4} \cmidrule(lr){5-7}
        $N$           & 85            &  71           & 42            & 85          &  71          & 42           \\
        mean $\pm$ SE & $13.40\pm0.67$& $14.03\pm0.71$& $15.51\pm0.87$&$7.53\pm0.55$& $8.02\pm0.58$&$10.23\pm0.68$\\
        25\%          &  9.55         & 9.74          & 10.86         &  5.59       &  6.69        &  7.57        \\
        median        &  12.30        & 12.75         & 14.84         &  8.28       &  8.96        & 11.58        \\
        75\%          &  15.26        & 16.82         & 19.41         & 10.57       & 11.08        & 13.42        \\ \midrule
        \multicolumn{7}{l}{Post-perihelion global curves} \\ \midrule
                      & \multicolumn{3}{c}{$k_1$ statistics}          & \multicolumn{3}{c}{$m_1$ statistics}        \\ \cmidrule(lr){2-4} \cmidrule(lr){5-7}
                      & new           & intermediate  & old           & new          & intermediate & old           \\ \cmidrule(lr){2-4} \cmidrule(lr){5-7}
        $N$           &  54           &  58           & 36            & 54           &  58          &  36           \\
        mean $\pm$ SE & $13.33\pm1.22$& $13.96\pm1.07$& $13.31\pm0.85$& $7.33\pm1.15$&$8.93\pm 0.54$& $11.35\pm0.54$\\
        25\%          & 9.35          & 10.03         & 9.53          &  6.90        &  6.79        &  9.68         \\
        median        & 11.54         & 12.66         & 13.21         &  8.75        &  9.71        & 11.57         \\
        75\%          & 14.41         & 15.38         & 16.03         & 10.48        & 11.51        & 13.43         \\ \bottomrule
        \vspace{1pt}
        \end{tabular}
        \tablefoot{Statistics of global curve $k_1$ and $m_1$ samples grouped by Oort group (new, intermediate and old) and orbital arc (pre- and post-perihelion). $N$ is the number of comets. Mean, standard error on the mean (SE), median and 25\% and 75\% quartiles are in units of mag/$\log{\text{au}}$.}
    \end{table*}

    \subsection{Comparing brightening using global curves}
    
    Analyzing the $k_1$ distributions for global curves reveals a similar trend, though the differences between the Oort groups are less pronounced (see Figure~\ref{fig:ecdf_k1_global}, and Tables~\ref{tab:pairwise_prob_kr_mr_strands} and \ref{tab:sample_stats_k1_m1_global_curves}). The smallest $p$-value occurs when comparing new and old comets ($p_\text{val}\approx0.03$). New and intermediate comets are statistically indistinguishable ($p_\text{val}\approx 0.45$), as are intermediate and old comets ($p_\text{val}\approx 0.23$). New comets have the lowest brightening slope (median $k_1=12.3$, interquartile range $\text{IQR}=5.7$), followed by intermediate comets (median $k_1=12.8$, $\text{IQR}=7.1$) and old comets (median $k_1=14.8$, $\text{IQR}=8.6$). The reduced distinction among global curves could be attributed to the smaller sample sizes; however, as discussed below, it may also be caused by trying to capture the brightening with a single set of parameters for all heliocentric distances.

    We note here that the brightening parameter \texttt{K1}, easily accessible from the SBDB, already hints at the same differences in behavior. New comets have the narrowest distribution and the lowest median $\texttt{K1}=6.5$, with an interquartile range (IQR) between 4.6 and 8.9. Old comets have the broadest distribution and the steepest brightening (median $\texttt{K1}=9.5$, IQR $[6.4,14.3]$) and intermediate comets have intermediate brightening (median $\texttt{K1}=7.8$, IQR $[5.4,10.3]$). If we take all three dynamical groups together, \texttt{K1} has a median of 7.5, with IQR $[5.3, 10.3]$.

    \begin{figure*}
    \centering
        \includegraphics[width=17cm]{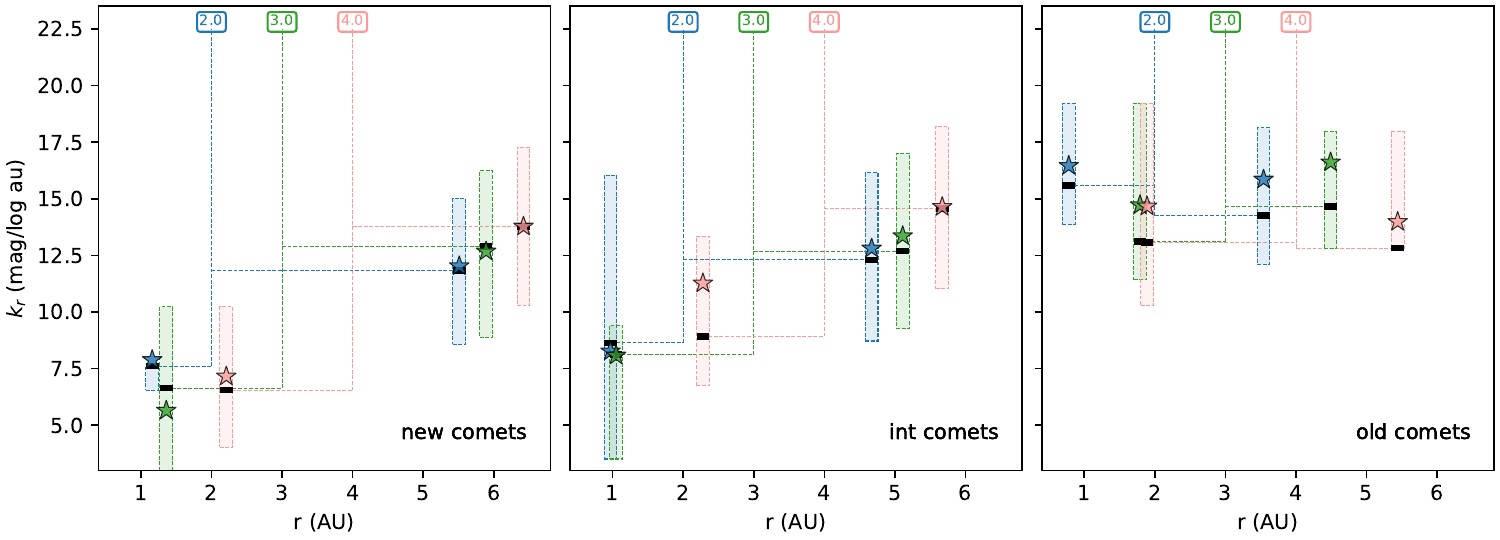}
            \caption{Brightening slope of strands as a function of heliocentric distance. A range of distance boundaries, shown as vertical dotted lines labeled by distance, separate strands into interior and exterior to the boundary. Two horizontal lines extending from each boundary to the left and to the right connect to boxes that correspond to interior and exterior strands, respectively. The $x$ coordinate of each box is the median of the midpoint in $\log{r}$ of the strands. The $y$ extent of each box contains the IQR of the $k_r$ distribution. Inside each box, a horizontal line marks the median and a star symbol indicates the mean of the $k_r$ distribution.}
            \label{fig:kr_vs_r}
    \end{figure*}

    \subsection{Evolution of brightening slope with heliocentric distance}

    To evaluate potential changes in brightening behavior with heliocentric distance, we analyzed separately strands that lie entirely within (interior) or beyond (exterior) a specific heliocentric distance. As before, we used debiased strand samples, ensuring that each distance range and each comet contribute the same number of strands. Figure~\ref{fig:kr_vs_r} illustrates the contrast between the $k_r$ distributions of interior and exterior strands for a range of distance boundaries.  

    Focusing on the boundary at 3 au, we find that new comets have significantly different $k_r$ medians inside and outside that distance (6.7 vs.~12.8, respectively). According to the Wilcoxon-Mann-Whitney U test, the null hypothesis that the medians are equal is rejected with $p_\text{val}\sim 10^{-4}$. For intermediate comets, the difference is notable but smaller (medians 7.8 vs.\ 12.5, $p_\text{val}=0.022$), and does not reach the $3\sigma$ significance threshold. Furthermore, the range of boundaries in Figure~\ref{fig:kr_vs_r} suggests that both groups experience a steady decline in brightening rate, from $\gtrsim$12.5 to less than 10 mag/log(au), as they approach the Sun. In contrast, the brightening slope of old comets shows no significant difference inside and outside 3 au (medians 13.4 vs.\ 14.6, $p_\text{val}=0.45$) and exhibits larger scatter. We note that using a smaller but more homogeneous sample of 21 LPCs, \citet{2024PSJ.....5..273Holt} also find a decreasing rate of brightening as comets approach the Sun. Their sample is dominated by new comets, with only 3 returning LPCs. 

    Additionally, both Figure~\ref{fig:kr_vs_r} and the median values suggest that the brightening behavior of newer and old comets differs primarily within 3 au. The U test confirms this, rejecting the null hypothesis when comparing the medians of new and old comets inside 3 au ($p_\text{val}\sim10^{-6}$). The difference between intermediate and old comets is less pronounced ($p_\text{val}=0.02$), falling short of the $3\sigma$ threshold. Outside 3 au, the same test indicates that all three groups have statistically indistinguishable medians ($p_\text{val}\geq0.27$).

    \subsection{Testing non-linear brightening models}
    
    If indeed the slopes of strands vary with heliocentric distance, global curves are unlikely to follow a strictly linear relationship as prescribed in Eq.~\ref{eq:heliocentric_magnitude}. To test this, we compared the original linear fit with a quadratic fit in $\log{r}$ using OLS fitting and the Akaike Information Criterion (AIC), as implemented in Python's \texttt{statsmodels} package. AIC accounts for both the goodness of fit (through the likelihood function) and model complexity (via a penalty for additional parameters). As anticipated, 83.8\% of global curves are better represented by a quadratic model. Interestingly, strands are also more accurately described by a quadratic fit: 68.9\% favor a quadratic model, while 31.1\% are better explained by a linear fit. As expected, the difference is less pronounced, given that strands cover narrower heliocentric ranges. Further analysis of the fit residuals (normality, mean, skew and homoscedasticity) for the two models confirmed the conclusion offered by AIC. Improving the modeling of secular brightening requires a more uniform sample, for instance focusing on data from the best characterized surveys. \citet{2024EPSC...17..324Snodgrass} found evidence for non-linear brightening in LPCs and propose an alternative empirical model where the rate of brightening varies linearly with heliocentric distance.

    \begin{figure}
        \resizebox{\hsize}{!}{\includegraphics{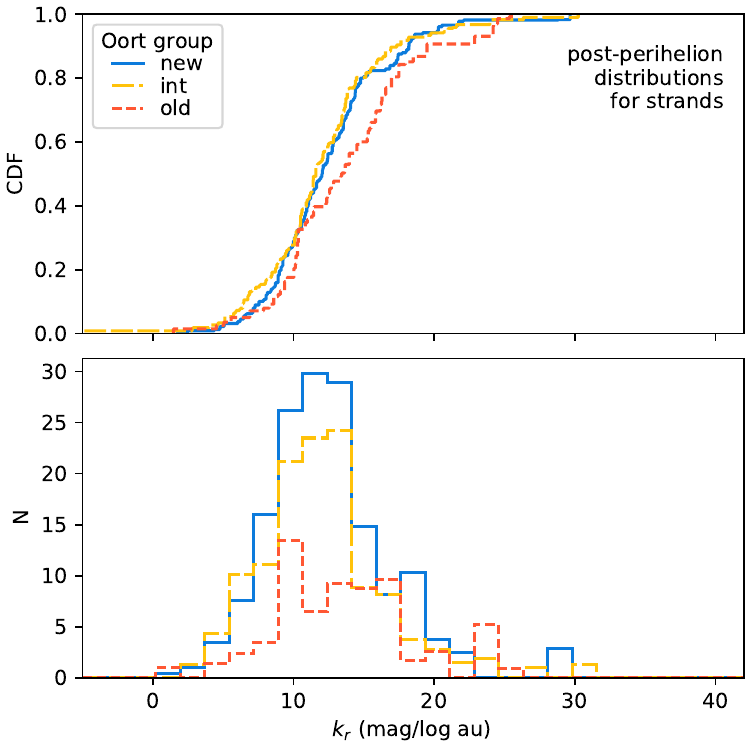}}
        \caption{Same as Figure~\ref{fig:ecdf_kr_strands} but for the $k_r$ fading parameter corresponding to post-perihelion strands.}
         \label{fig:ecdf_kr_strands_out}
    \end{figure}

    \begin{figure}
        \resizebox{\hsize}{!}{\includegraphics{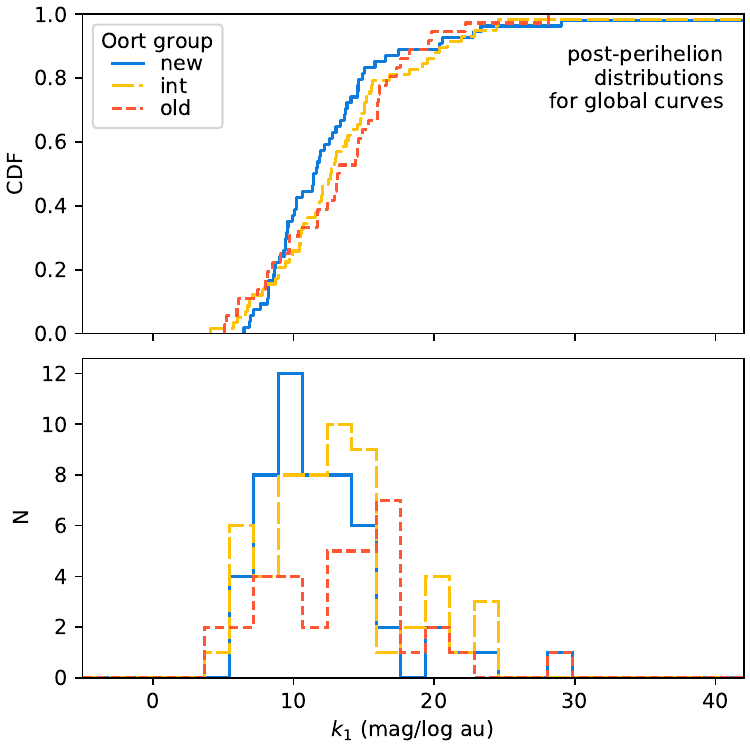}}
        \caption{Same as Figure~\ref{fig:ecdf_k1_global} but for the $k_1$ fading parameter corresponding to post-perihelion global curves.}
        \label{fig:ecdf_k1_global_out}
    \end{figure}

    \begin{figure}
        \resizebox{\hsize}{!}{\includegraphics{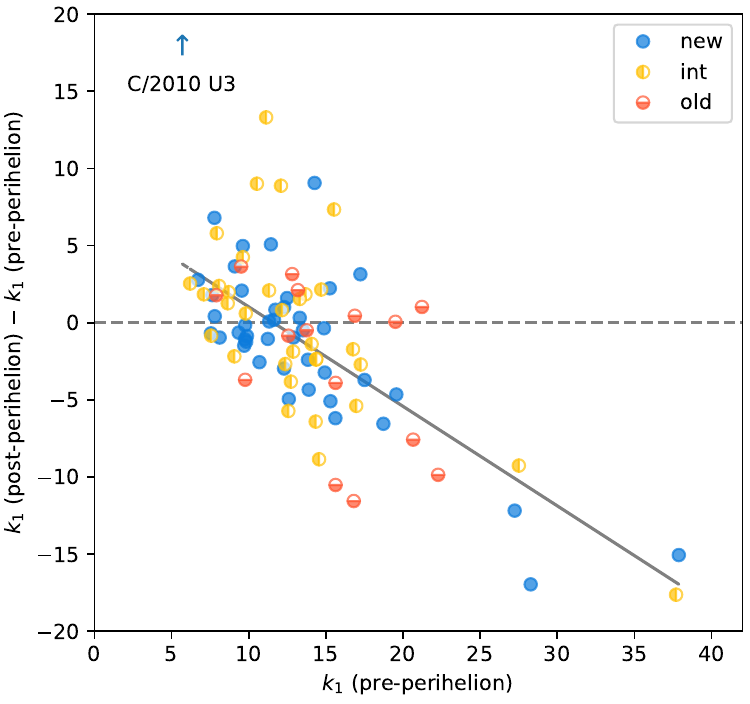}}
        \caption{Difference between brightening and fading slopes of global curves. A robust linear model fit of the form $\Delta k_1=-0.65(k_1-11.6)$ is shown (adjusted $R^2=0.27$).}
        \label{fig:incoming_vs_outgoing}
    \end{figure}
    
    \begin{table}
        \caption{Brightening/fading slope change statistics
        \label{tab:k1_in_vs_out}}
        \centering
        \begin{tabular}{lrrrr}
        \toprule
        $\Delta k_1$ statistic & new & intermediate & old & all comets \\
        \midrule
        $N$ & 44 & 33 & 15 & 92 \\
        > 0 & 18 & 17 & 7 & 42 \\
        < 0 & 26 & 16 & 8 & 50 \\
        minimum & $-17.0$ & $-17.6$ & $-11.6$ & $-17.6$\\
        25\% & $-3.0$ & $-2.7$ & $-5.7$ & $-3.3$\\
        median & $-0.7$ & 0.6 & $-0.5$ & $-0.6$\\
        75\% & 1.6 & 2.2 & 1.4 & 2.0\\
        maximum & 63.5 & 13.3 & 3.6 & 63.5\\
        Spearman $r$ & $-0.57$ & $-0.57$ & $-0.46$ & $-0.57$ \\
        \bottomrule
        \end{tabular}
        \tablefoot{Statistics of $\Delta k_1$, which measures change in $k_1$ between pre- and post-perihelion orbit. Columns show results per Oort group and for all comets. Rows indicate sample size, number of comets with positive and negative $\Delta k_1$, minimum, quartiles and maximum of $\Delta k_1$ distributions, and Spearmank rank coefficient of the correlation between $\Delta k_1$ and $k_1$ pre-perihelion (see Figure~\ref{fig:incoming_vs_outgoing}.)}
        \vspace{1pt}
    \end{table}

    \subsection{Post-perihelion fading behavior}
    
    If we look at the post-perihelion fading behavior, we find that it is more consistent among the different Oort groups (see Figures~\ref{fig:ecdf_kr_strands_out} and \ref{fig:ecdf_k1_global_out} and Tables~\ref{tab:sample_stats_kr_mr_strands} and \ref{tab:sample_stats_k1_m1_global_curves}). Parameter $k_r$ (now a fading slope) for outgoing strands of newer comets tends to be slightly shallower than that of old comets; however, the distributions are not significantly different (see Table~\ref{tab:pairwise_prob_kr_mr_strands}). Moreover, the fading slope samples of outgoing global curves of new, intermediate and old comets are statistically indistinguishable (Table~\ref{tab:sample_stats_k1_m1_global_curves}). 
    
    For the 92 comets with both incoming and outgoing data, Figure~\ref{fig:incoming_vs_outgoing} compares the global curve pre- and post-perihelion, illustrating how the $k_1$ slope changes between brightening and fading (see also Table~\ref{tab:k1_in_vs_out}). The median change in slope is $\Delta k_1 = -0.6$ mag/log(au) and the number of comets that increase or decrease $k_r$ after perihelion is roughly equal. Half of comets exhibit slope changes within $-3.3<\Delta k_1<2.0$ mag/log(au). Figure~\ref{fig:incoming_vs_outgoing} suggests a correlation between $\Delta k_1$ and incoming $k_1$, particularly for newer comets. A robust linear fit yields the relation $\Delta k_1=-0.65(k_1-11.6)$. If the post-perihelion fading slope is related to the subsequent pre-perihelion brightening, the apparent trend would lead to $k_1$ converging toward a value near 11.6 mag/log(au) after repeated perihelion visits.

    \begin{figure}
        \resizebox{\hsize}{!}{\includegraphics{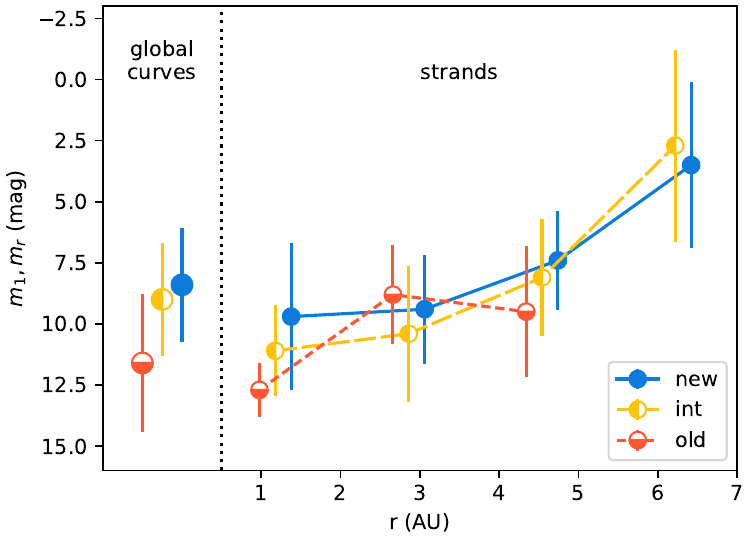}}
        \caption{Median $m_r$ for strands as a function of heliocentric distance and Oort group are shown to the right of the vertical dashed line. Strands are binned in equal-sized $r_\text{min}$ bins (linear $r$) and the median $m_r$ and median absolute deviation (MAD) are plotted as points and error bars, respectively. Bin centers coincide with the $x$-axis location of the "int." symbols and the "new" and "old" symbols where horizontally nudged for clarity. To the left of the vertical dashed line are the $m_1$ distribution median plus or minus MAD for global curves.}
        \label{fig:m1_mr_vs_rmin}
    \end{figure}

    \begin{figure}
        \resizebox{\hsize}{!}{\includegraphics{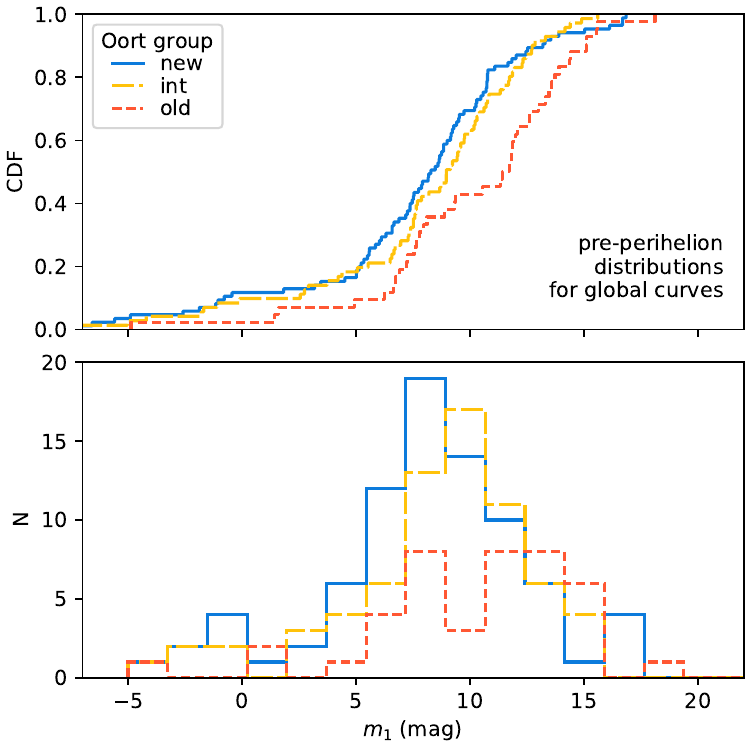}}
        \caption{Empirical cumulative probability distribution of the $m_1$ parameter for pre-perihelion global curves, by Oort group.}
        \label{fig:ecdf_m1_global_in}
    \end{figure}

    \subsection{Investigating the \texorpdfstring{$m_1$}{m1} parameter}
    \label{subsec:results_m1_parameter}

    As mentioned above, $m_r$ and $m_1$ are less reliably extracted from the MPC data due to inconsistencies in the photometric analysis by different observers. Furthermore, the aforementioned changes in $k_r$ as the comet approaches the Sun imply that $m_r$ fits to strands will depend on heliocentric distance. This is apparent in Figure~\ref{fig:m1_mr_vs_rmin}, where $m_r$ values increase as comets approach the Sun, reflecting the corresponding decrease in $k_r$. It is reassuring that the $m_1$ values obtained from fitting global curves (to the left of the dashed line in the Figure) are consistent with the values of $k_r$ for strands approaching 1 au (right of dashed line). With these caveats in mind, we compare in Figure~\ref{fig:ecdf_m1_global_in} the distributions of the $m_1$ parameter for new, intermediate, and old comets. New comets exhibit brighter median $m_1$ values than old comets (Table~\ref{tab:sample_stats_k1_m1_global_curves}), and the respective distributions are statistically incompatible according to the bootstrap test ($p_\text{val}=0.0015$, Table~\ref{tab:pairwise_prob_kr_mr_strands}). The $m_1$ distributions for intermediate and old comets are not incompatible at the $3\sigma$ level ($p_\text{val}=0.0135$), and new and intermediate comets are indistinguishable ($p_\text{val}=0.5$). In Subsection~\ref{subsec:brightness_of_new_comets} we discuss the possibility that the brighter median $m_1$ for new comets when compared to old comets is the result of a selection bias.

    \begin{figure}
        \resizebox{\hsize}{!}{\includegraphics[width=\hsize]{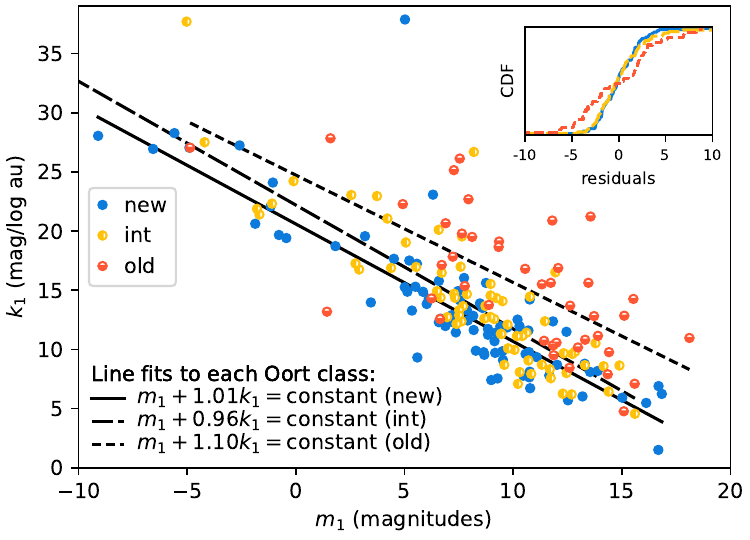}}
        \caption{Relation between $m_1$ and $k_1$ for pre-perihelion global curves, color-coded by Oort group. Points correspond to individual comets. Black lines are RLM fits to each Oort group. The respective fit coefficients of determination are $R^2_\text{new}=0.71$, $R^2_\text{int}=0.71$, and $R^2_\text{old}=0.45$.}
        \label{fig:m1_vs_k1}
    \end{figure}

    \subsection{The link between the brightening parameters}
    \label{subsec:m1_k1_link}
    
    The link between $m_r$ and $k_r$ for strands is straightforward since they are derived from local fits. However, it is interesting to examine the relationship between $m_1$ and $k_1$ for global curves, as it captures secular trends in brightening behavior. Figure~\ref{fig:m1_vs_k1} illustrates this relation for pre-perihelion global curves, grouped by Oort group. Linear fits are also shown, together with analytical expressions and coefficients of determination, $R^2$. The slopes of these fits are close to $-1$, particularly for new comets. The fit is tightest for new comets, yielding residuals with a median of $-0.1$ and an IQR of $[-1.6, 1.6]$, and intermediate comets (median: $-0.2$, IQR: $[-1.7, 1.8]$). For old comets the residuals become more dispersed (median: $-0.5$, IQR: $[-3.2, 2.6]$). To assess whether the dispersion of residuals differs between dynamical groups, we applied the Fligner-Killeen test, a non-parametric method robust to non-normality that ranks absolute deviations from the median. We compared the residuals of new vs. old and intermediate vs. old comets, testing the null hypothesis that their dispersions are equal. The test yielded $p_\text{val}\sim 10^{-5}$ for new vs.~old, indicating a highly significant difference. For intermediate vs.~old, we obtained $p_\text{val}=0.00145$, which is still significant at $3\sigma$ significance. In the case of new comets, the approximate constancy of $m_1 + 1.01 k_1$ suggests that these comets have similar magnitudes at $r\approx10^{1.01} \approx 10.2$ au. This uniformity of behavior is intriguing, and may be influenced by a selection bias. However, while the absence of new comets with fainter $m_1$ can be explained in this context, the lack of examples with steeper $k_1$ values remains unexplained. A more detailed analysis of this trend is beyond the scope of this paper, but determining its underlying cause will likely require a well-characterized survey such as LSST.

    \begin{figure}
        \resizebox{\hsize}{!}{\includegraphics{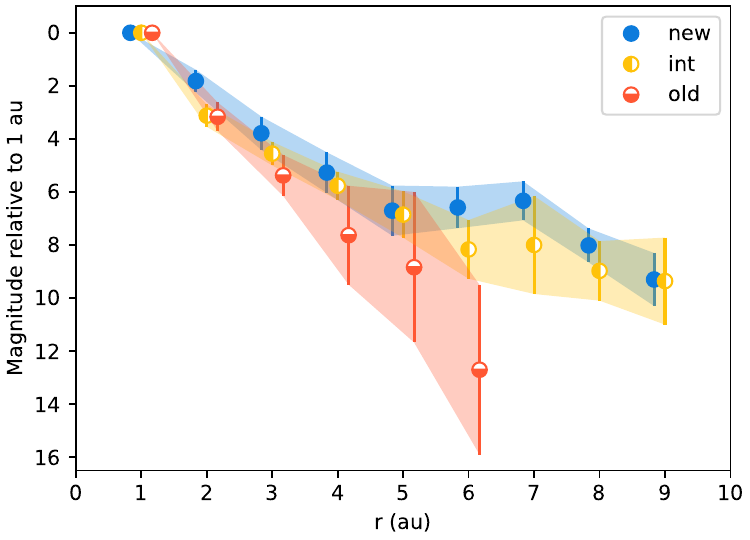}}
        \caption{Average relative magnitude brightening per Oort group. Median magnitudes are calculated for each comet in heliocentric distance bins, grouped by Oort classification and averaged. See text for details. Shading connecting points and $1\sigma$ standard error bars are added to guide the eye. Bin centers coincide with the $x$-axis location of the "int." symbols and the "new" and "old" symbols where horizontally nudged for clarity.}
        \label{fig:delta_mag_vs_r}
    \end{figure}

    \begin{table*}
        \caption{Magnitude brightening distribution as a function of heliocentric distance, per Oort group.}
        \label{tab:brightening_distributions}
        \centering
        \begin{tabular}{llrrrrrrrr}
        \toprule
                                &  & \multicolumn{8}{c}{brightening in magnitudes from $r$ to 1 au} \\ \cmidrule(lr){3-10}
        Oort group              & $r$ (au) &$\mu$ &$\sigma$& 10\% & 25\% & 50\% & 75\% & 90\%  & $N$ \\ \midrule
        \multirow[t]{9}{*}{new} & 1 & 0.00 & 0.00 & 0.00 & 0.00 & 0.00 & 0.00 &  0.00 & 13  \\
                                & 2 & 1.82 & 1.48 & 0.24 & 0.67 & 1.68 & 2.45 &  3.26 & 12  \\
                                & 3 & 3.79 & 2.09 & 1.53 & 2.38 & 3.70 & 5.22 &  5.85 & 11  \\
                                & 4 & 5.27 & 2.42 & 2.81 & 3.20 & 5.32 & 7.00 &  8.54 & 10  \\
                                & 5 & 6.70 & 2.68 & 3.53 & 4.80 & 6.40 & 8.95 &  9.90 &  8  \\
                                & 6 & 6.58 & 1.92 & 4.68 & 5.26 & 6.73 & 6.80 &  8.35 &  6  \\
                                & 7 & 6.33 & 1.47 & 5.12 & 5.14 & 6.07 & 7.26 &  7.77 &  4  \\
                                & 8 & 8.02 & 1.28 & 6.79 & 7.38 & 8.13 & 8.77 &  9.15 &  4  \\
                                & 9 & 9.30 & 1.41 & 8.50 & 8.80 & 9.30 & 9.80 & 10.10 &  2  \\ \midrule
        \multirow[t]{9}{*}{intermediate} & 1 & 0.00 & 0.00 & 0.00 & 0.00 & 0.00 & 0.00 & 0.00 & 16  \\
                                & 2 & 3.12 & 1.57 & 1.36 & 1.66 & 3.25 & 4.51 &  5.07 & 14  \\
                                & 3 & 4.56 & 1.54 & 2.62 & 3.64 & 4.85 & 5.33 &  5.40 & 12  \\
                                & 4 & 5.77 & 1.77 & 3.80 & 4.47 & 5.80 & 6.80 &  7.65 & 11  \\
                                & 5 & 6.85 & 1.95 & 4.76 & 6.41 & 7.60 & 8.10 &  8.34 &  5  \\
                                & 6 & 8.16 & 2.47 & 5.77 & 7.50 & 8.20 & 9.10 & 10.49 &  5  \\
                                & 7 & 8.00 & 2.61 & 6.52 & 7.07 & 8.00 & 8.92 &  9.48 &  2  \\
                                & 8 & 8.97 & 1.94 & 7.36 & 8.16 & 9.50 &10.05 & 10.38 &  3  \\
                                & 9 & 9.36 & 2.32 & 8.05 & 8.54 & 9.36 &10.18 & 10.67 &  2  \\ \midrule
        \multirow[t]{9}{*}{old} & 1 & 0.00 & 0.00 & 0.00 & 0.00 & 0.00 & 0.00 &  0.00 & 15  \\
                                & 2 & 3.17 & 1.95 & 0.80 & 1.32 & 3.90 & 4.50 &  4.68 & 13  \\
                                & 3 & 5.37 & 2.44 & 2.86 & 3.89 & 5.25 & 6.33 &  8.70 & 10  \\
                                & 4 & 7.63 & 4.17 & 4.03 & 4.30 & 6.22 &10.18 & 12.23 &  5  \\
                                & 5 & 8.84 & 4.91 & 5.82 & 6.00 & 6.31 &10.40 & 12.86 &  3  \\
                                & 6 & 2.70 & 4.53 &10.14 &11.10 &12.70 &14.30 & 15.26 &  2  \\
        \bottomrule
        \end{tabular}
        \tablefoot{Each row lists average and standard deviation, plus quantiles (based on sample of size $N$) of how many magnitudes a comet will brighten from a given heliocentric distance to 1 au. See text for details.}
        \vspace{1pt}
    \end{table*}

    \subsection{Brightening trends from direct magnitude analysis}

    As an alternative to modeling the reported magnitudes using Eq.~\eqref{eq:heliocentric_magnitude}, we can directly analyze the magnitudes to identify trends in brightening behavior. Figure~\ref{fig:delta_mag_vs_r} shows how the magnitude of incoming comets, relative to their magnitude at 1 au, varies with heliocentric distance. To generate these curves, we consider only comets that reach 1 au and divide each comet's measurements into heliocentric distance bins (1 au wide and 1 au apart, centered at $1, 2, 3, ...$ au.). For each bin, we compute the median magnitude and subtract the median magnitude at 1 au to produce a relative brightening curve for each comet. We then derive average brightening curves for each Oort group by computing the mean and standard deviation in each bin. The resulting values, along with quantiles of the brightening distribution in each bin, are listed in Table~\ref{tab:brightening_distributions}. These results confirm that newer comets brighten more gradually as they approach the Sun, while old comet display more scatter and no data beyond $r\approx 6$ au. As a general trend, inbound new and intermediate comets brighten by approximately 10 magnitudes between 10 au and 1 au. We note that these brightening trends are based on a limited number of comets (see Table~\ref{tab:brightening_distributions}).

    \subsection{Trends with other orbital elements}
    \label{subsec:trends_with_orbital_elements}

    The orbital inclination distributions of new, intermediate and old comets in our sample are similar and all uniform between 0 and 180 degrees. Likewise, orbital inclination does not seem to influence the $k_1$ distribution: no trends are observed between the two, either overall or within individual Oort groups. However, perihelion distance and Oort group are linked in our sample: new comets have a larger median perihelion distance ($q_\text{new} = 3.1$ au) compared to intermediate comets ($q_\text{int.} = 2.8$ au) and old comets ($q_\text{old} = 1.8$ au). 
    
    \citet{1976NASSP.393..410Meisel} noted that splitting comets by perihelion distance results in a stronger distinction than using $1/a_0$. Specifically, they divided comets into two groups: those with perihelion distances inside and outside $q = 1.25$ au. Their choice of threshold was partly constrained by the limited number of comets in their sample with $q > 3$ au and the absence of any with $q > 3.8$ au. Given that our dataset includes a larger number of high-perihelion comets, we investigated whether splitting the sample at a larger $q$ would yield a greater difference in brightening behavior. Using the bootstrap procedure described at the start of this section, we tested $q_\text{split}$ from 1 to 4 au in steps of 0.25 and found no significant difference between $k_r$ interior and exterior to any value of $q_\text{split}$.

    \section{Discussion}

    We found evidence that new comets behave differently from returning comets as they approach the Sun. The main differences are that newer comets are typically brighter at all distances (Figures~\ref{fig:ecdf_m1_global_in} and \ref{fig:delta_mag_vs_r}), brighten more slowly, particulary inside 3 au (Figures~\ref{fig:ecdf_kr_strands} and \ref{fig:kr_vs_r}), and display 
    a tighter correlation between $m_1$ and $k_1$ (Figure~\ref{fig:m1_vs_k1}) than old comets. Intermediate comets display intermediate behavior, which tends to be closer to that of new comets. The different distributions for the brightening slope are most obvious when calculated for individual strands ($k_r$) which capture the effects of changing slope with heliocentric distance (Figure~\ref{fig:ecdf_kr_strands}.) We discuss these points below.

    \subsection{On the different brightening of new and returning comets}
    
    A number of authors have found that new comets show a slower increase in brightness as they approach the Sun \citep{1951BAN....11..259Oort, 1976NASSP.393..410Meisel, 1978M&P....18..343Whipple, 1995ICQ....17..168Green}. Our analysis uses a much larger set of comets and essentially confirms this result, while adding that the brightening slope of new comets decreases from $k_r\approx13$ mag/log(au) beyond 3 au to $k_r\approx7$ mag/log(au) inside 3 au, where it plateaus (see 
    Figure~\ref{fig:kr_vs_r}). Converting Eq.~\ref{eq:heliocentric_magnitude} to total flux relative to 1 au, these translate into $r^{-k_r/2.5}$, i.e., an approximately $r^{-3}$ dependence inside 3 au, which rises to approximately $r^{-5}$ and even steeper beyond that distance. This slowing rate of brightening is corroborated by a recent dedicated observing campaign, which collected data on 21 LPCs and using the Las Cumbres Observatory network of telescopes \citep{2024PSJ.....5..273Holt}. Despite being smaller in number, their observations were calibrated and processed in a uniform manner, resulting in high quality data that also show that LPC secular brightening is not adequately described by Eq.~\eqref{eq:heliocentric_magnitude} with a single set of parameters from beyond Jupiter all the way to perihelion. Even though the authors report no difference between the behavior of new and returning comets, their sample included only 3 returning comets, so we interpret it as representative of new comets.
    
    Qualitatively, the behavior of newer comets fits with a proposed explanation \citep{1978M&P....18..343Whipple} that the availability of hypervolatile ices driving their activity is exhausted before they reach perihelion because those ices are concentrated on a thin, near-surface, ``frosting'' layer developed by cosmic ray irradiation in the Oort cloud over the age of the solar system. The release of ices in order of decreasing volatility \citep[e.g.,][]{1982come.coll...85Delsemme,2004come.book..317Meech} results in a decreasing rate of brightening, ending with the least volatile available ice, of which the value $k_r\approx 7.5$ mag/log(au) is representative (see Figure~\ref{fig:kr_vs_r}). Repeated perihelion visits lead to progressive build-up of a patchy mantle of refractory material mixed with low volatility ices. As the fraction of the surface covered by this mantle increases, irregularities in thickness and spatial distribution may explain the larger, more scattered and less $r$-dependent $k_r$ of old comets.

    New and intermediate comets behave similarly inside 3 au, but new comets brighten more slowly than even intermediate comets at distances 4--7 au. This suggests that the exhaustion or severe depletion of hypervolatiles in new comets occurs further out, consistent with activity starting at large distances \citep{2021AJ....161..188Jewitt}, even though CO has been detected in \object{C/2017 K2} at 6.7 au \citep{2021ApJ...914L..17Yang} and CO, CH$_3$ and CH$_3$OH have been detected in \object{C/2007 N3} at 1.3 au \citep{2016Icar..278..301DelloRusso}, both dynamically new comets. Detailed observations and modeling of the activity of the well-studied \object{C/2017 K2} exemplify the intricacies of decoding the interplay of different physical mechanisms \citep[e.g.,][]{2019AJ....157...65Jewitt,2020A&A...636L...3Fulle,2022ApJ...924...37Bouziani}. Such an effort is beyond the scope of this paper.
    
    The differences between the photometric behavior of new and returning comets have been highlighted as a possible explanation for the odd proportion of new to returning comets, in that the former is several times larger than would be expected from a semi-stationary state where orbits evolve toward smaller semi-major axis with repeated visits \citep{1950BAN....11...91Oort,1999Icar..137...84Wiegert}. This lack of returning comets is generally explained by strong ``fading'' of comets following their first return to the planetary system from the Oort cloud, which may begin beyond Neptune \citep{2022SciA....8M9130Kaib}. The progressive change in brightening behavior we observe from new, to intermediate, to old comets supports the idea of evolution in the direction of fading as a result of repeated perihelion visits. The subjacent cause of the fading is not identified with certainty and may be due to a combination of effects. For instance, activity driven torques may spin-up the nuclei of comets to rotational disruption \citep{2022AJ....164..158Jewitt}. This effect is strongly dependent on perihelion distance, which does diminish as comets evolve from dynamically new to old \citep[see Subsection~\ref{subsec:trends_with_orbital_elements}, and][]{2009Sci...325.1234Kaib}.

    \subsection{On the intrinsic brightness of new comets}
    \label{subsec:brightness_of_new_comets}

    A study of distant activity of 50 LPCs found that new comets display a higher level of activity than returning comets \citep{2016AJ....152..220Sarneczky}. This result, based on observations obtained at a single observatory and subject to uniform processing, is supported by our larger albeit less uniform study: we, too, find that new comets are generally brighter than intermediate comets, which are, in turn, brighter than old comets (Table~\ref{tab:sample_stats_k1_m1_global_curves} and Figures~\ref{fig:m1_mr_vs_rmin} and \ref{fig:ecdf_m1_global_in}).  A trend of decreasing intrinsic activity brightness with increasing dynamical ages suggests that it may be related to the cumulative effect of multiple perihelion passages. Indeed, SPCs tend to have fainter total absolute magnitudes and lower activity than LPCs as a whole \citep[e.g.,][]{2023MNRAS.526..246Betzler}. 
    
    Could our finding that new comets are brighter than old comets be the result of a selection bias? We believe this is unlikely. In our sample, new comets have a larger median perihelion distance than old comets ($q_\text{new} = 3.1$ au vs. $q_\text{old} = 1.8$ au). Although new comets would be expected to brighten by approximately 1.1 mag over this range (Table~\ref{tab:brightening_distributions}), the observed difference in median $m_1$ between new and old comets is 3.3 mag (Table~\ref{tab:sample_stats_k1_m1_global_curves}), suggesting that selection effects alone cannot account for the observed trend.
    
    The total brightness of an active comet is typically dominated by the coma, which scales with nucleus size assuming the same activity level per surface area \citep{1999A&A...352..327Fernandez, 2011MNRAS.416..767Sosa,2022AJ....164..158Jewitt}. So, one explanation for the observed trend is that new comets possess larger nuclei. This is expected, as comets lose mass due to sublimation and splitting, which affects new comets more frequently than returning ones, and will result in an overall decrease in average size for returning comets compared with new ones \citep{1980A&A....85..191Weissman}. Alternatively, dynamically new comets may appear brighter due to intrinsically higher activity levels than those of older, recurrent comets \citep[e.g.,][]{2014A&A...561A...6MazzottaEpifani}, possibly as a result of higher availability of surface volatiles \citep{2016AJ....152..220Sarneczky}.  The trend is also broadly consistent with the traditional concept of mantling. In this scenario, a non-volatile mantle of refractory material develops over successive perihelion passages, reducing sublimation and leading to progressively lower activity, eventually resulting in a dormant or extinct comet.  It is also possible that all these factors contribute to the observed brightness differences.

    To investigate the effect of nucleus size, we used two independent sources of data.  \citet{2024MNRAS.531..859Robinson} compiled a set of LPC nucleus size estimates, most of which were obtained using NEOWISE infrared data \citep{2017AJ....154...53Bauer} from the WISE mission \citep{2011ApJ...731...53Mainzer}. Of those, 18 overlap with our comet data, which we restricted to comets with known Oort group and inbound global curve parameters $k_1$ and $m_1$. For the overlapping LPCs, we find that new comets are generally larger (median $r=8.1$ km, $N=7$) than intermediate (median $r=4.8$ km, $N=6$) and old comets (median $r=4.0$ km, $N=5$).  \citet{2022AJ....164..158Jewitt} published a sample of LPC sizes derived from total production rates and from non-gravitational acceleration data. The two independent methods they used produce consistent size estimates, which are averaged to produce a list of comet nucleus sizes that partly intersects with our data. Here, too, we find that new comets are generally larger (median $r=1.6$ km, $N=6$) than intermediate (median $r=1.1$ km, $N=6$) and old comets (median $r=1.0$ km and $N=5$). Conveniently, this sample includes radii smaller than that of \citet{2017AJ....154...53Bauer}, allowing us to extend the scope of our analysis. Combining both sources for nucleus size, we retain the trend: new comets are larger (median $r=4.5$ km with IQR $1.7<r<8.4$ km), followed by intermediate comets (median $r=3.1$ km with IQR $1.7<r<5.0$ km) and old comets (median $r=2.5$ km with IQR $1.0<r<4.0$ km). It is interesting that intermediate comets lie closer to old comets, consistent with most mass loss occurring in the first few perihelion passages. We find no relation between inbound global curve brightening slope $k_1$ and nucleus radius. The Spearman Rank Correlation (SPR) test assigns a $p_\text{value}=0.854$ to the correlation. However, as expected, nucleus size does appear to correlate with the $m_1$ parameter (SPR test $p_\text{value}=0.017$). It is important to note that we are dealing with small samples, resulting in low statistical power, which lowers the chance to detect real features, and potential susceptibility to outliers affecting the results.
    
    To assess the influence of composition, we rely on \citet{2024MNRAS.531..859Robinson} who identified a significant log-log linear correlation between nucleus size and CO/H$_2$O coma abundance. The authors posit that this trend supports a semi-empirical model of near-surface volatile re-entrapment driven by early radiogenic heating \citep{2022MNRAS.514.3366Malamud}. This model proposes that larger nuclei, able to retain radiogenic heat, experienced volatile migration to outer layers, leading to a differentiated volatile distribution that sustains activity over multiple perihelion passages, particularly in younger comets. Smaller nuclei, with lower heat retention, would have limited, undifferentiated volatile reservoirs, leading to a stable but lower CO/H$_2$O ratio and a decline in activity with age. The model predicts that dynamically younger comets should be larger and display higher CO/H$_2$O abundances and greater activity.  Although Robinson's sample primarily includes SPCs, it contains some overlap with the LPCs studied here.  Using their composition data, we find that new comets show lower CO/H$_2$O abundances (median $\text{CO}/\text{H}_2\text{O}=0.10$, $N=3$) compared to intermediate (median $\text{CO}/\text{H}_2\text{O}=0.21$, $N=2$) and old comets (median $\text{CO}/\text{H}_2\text{O}=0.15$, $N=2$), which opposes the expected trend. Interestingly, \citet{2022PSJ.....3..247HPinto} find an increasing CO production with increasing $1/a_0$, i.e. with dynamical age. Their interpretation is that a 4.5 Gyr exposure to cosmic-rays in the Oort cloud leads to the erosion of all the near-surface CO, leaving almost none available to sublimate during a first perihelion passage \citep{2020ApJ...901..136Maggiolo}. Return passages will increasingly expose CO that was buried, increasing its relative importance to activity.

    \begin{figure}
        \resizebox{\hsize}{!}{\includegraphics{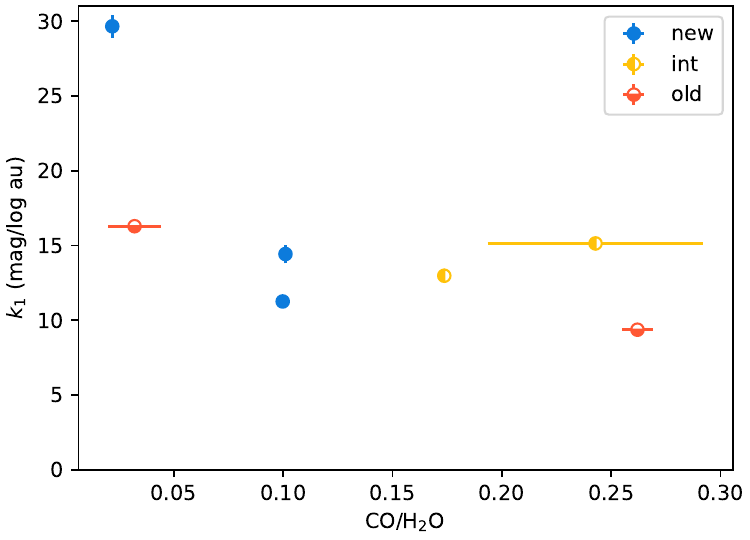}}
        \caption{Relation between the brightening slope $k_1$ of global curves and CO/H$_2$O abundance in the coma.}
        \label{fig:k1_vs_composition}
    \end{figure}

    We also examined the potential relationship between CO/H$_2$O abundance and brightening slope. To explore this, we analyzed the $k_r$ values of strands nearest in time to the gas abundances measurements. Figure~\ref{fig:k1_vs_composition} illustrates this relationship, suggesting that higher CO/H$_2$O abundances tend to correspond with smaller brightening slopes. However, the SPR test yielded a $p_\text{val}=0.09$, indicating a lack of statistical significance. We reiterate the caution above, that this discussion is based on very small numbers. More observations are needed to test model predictions in a compelling manner.
    
    Despite the small sample size, these findings suggest that the higher intrinsic brightness of dynamically new comets is more likely associated with larger nucleus size. We tentatively refute \citet{2022MNRAS.514.3366Malamud}'s prediction regarding hypervolatile abundances and dynamical age, but this may be due to small-sample statistics. However, we note that \citet{1995Icar..118..223AHearn} similarly found no dependence of composition on dynamical age, nor evidence for differentiated nuclei. The thermal processing of the layer contributing to the observed activity needs to be further investigated as discussed below.

    \subsection{On the uniform behavior in new comets}

    Figure~\ref{fig:m1_vs_k1} suggests that new comets display tighter correlations between $m_1$ and $k_1$ than old comets. This is supported by the linear fit $R^2$ values listed in the caption, and by analysis of the residuals (see Subsection~\ref{subsec:m1_k1_link}). The brightening parameters for global curves of new comets are such that $m_1+1.01k_1\sim\text{constant}$. Such a relationship would imply that these comets have a similar magnitude around 10 au. One possibility is that this is the result of a discovery bias, where known new comets are the brightest members of the underlying population. However, this does not readily explain the paucity of new comets with $k_1$ values steeper than those lying on the $m_1$-$k_1$ relation (see Figure~\ref{fig:m1_vs_k1}). If not due to a selection bias, the simplest explanation for this uniformity is that the layer contributing to the activity of dynamically new comets is similar in active area (or size) and composition as they enter the region of giant planets. Figure~\ref{fig:delta_mag_vs_r}, which shows the average magnitude of comets at different distances relative to 1 au, qualitatively supports the notion that new comets display more uniform behavior than old comets.

    The regular behavior of dynamically new comets has been noted previously. In their observational study of 50 LPCs beyond 5.2 au \citet{2016AJ....152..220Sarneczky} found that dynamically new comets tend to exhibit more regular and smoothly evolving dust production, with more symmetrical comae, suggesting a more isotropic outflow compared to returning comets. They also observed that dynamically young comets are intrinsically brighter, displaying dust activity at large distances, which they interpret as an indication of uniformity in the initial composition and activity of new comets.

    It is plausible that nuclei sent to the Oort cloud early in the solar system's evolution, now returning for the first time, underwent uniform processing distinct from that of their dynamically evolved counterparts. If these nuclei also had similar initial compositions and sizes, they might behave more consistently upon their first return to the inner solar system. However, it remains difficult to explain how spin-axis obliquity, an important factor in controlling activity \citep{1978M&P....18..343Whipple,2019A&A...623A.120Marschall}, could be made uniform. This likely remains a source of variability in activity patterns between comets. Since spin-axis orientation is believed to cause strong pre-/post-perihelion activity asymmetries \citep{2019A&A...623A.120Marschall}, the fact that we find roughly equal numbers of comets with decreasing and increasing brightening rates past perihelion argues for randomly oriented spin axes.

    Early thermal processing of LPC nuclei before ejection to the Oort Cloud may also introduce variation. \citet{2024PSJ.....5..243Gkotsinas} have investigated the thermal history of planetesimals that formed roughly 20 to 30 au from the Sun and were dynamically ejected to the Kuiper belt, scattered disk and the Oort cloud during the early phases of the solar system. They conclude that Oort cloud planetesimals are the least thermally processed population, with a greater chance of retaining their original ice content. This is attributed to the relatively rapid outward scattering they experience, primarily due to interactions with Jupiter and Saturn. This swift outward movement limits their exposure to high temperatures in the inner solar system. The simulations show that around 60\% of Oort cloud planetesimals retain a portion of their initial CO ice.
    
    The authors suggest that comets implanted in the Oort cloud at different heliocentric distances have statistically followed different orbital trajectories which could explain the observed differences in their volatile and hypervolatile gas production rates. Comets implanted beyond the Oort spike (typically around 10,000 au) are considered dynamically new, while comets implanted at closer distances are considered dynamically old. Their simulations show that dynamically new comets are more processed in terms of CO content compared to dynamically old comets, which could explain why observations suggest that dynamically new comets produce more CO$_2$ than CO, while dynamically old comets appear to be more CO-dominant \citep{1995Icar..118..223AHearn, 2022PSJ.....3..247HPinto}. Figure~\ref{fig:k1_vs_composition} supports the idea that dynamically new comets have lower CO/H$_2$O abundance, even though the sample is small.

    \begin{figure}
        \resizebox{\hsize}{!}{\includegraphics{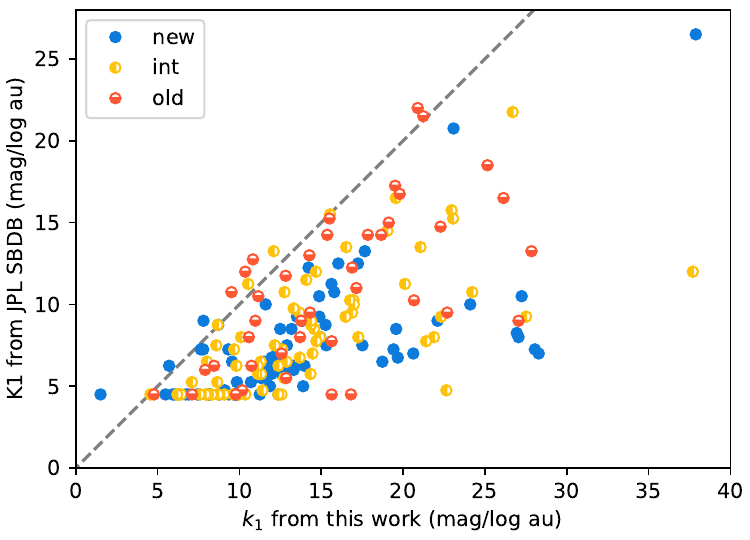}}
        \caption{Comparison of brightening slope \texttt{K1} from JPL SBDB with $k_1$ derived in this paper. Both axes are in units of mag/log(au), and a dashed gray line indicates $\texttt{K1} = k_1$.}
        \label{fig:K1_vs_k1}
    \end{figure}

    \subsection{Comparing JPL SBDB and paper-derived parameters}
    \label{subsec:jplvspaper}
    
    JPL SBDB provides parameters \texttt{M1} and \texttt{K1} (see Eq.~\ref{eq:total_magnitude}) for most comets. As noted earlier, these parameters are derived by fitting all observations submitted to the MPC, using a less stringent approach than the one employed in this paper. In particular, the fitting procedure combines both pre- and post-perihelion data for each comet, producing a single \texttt{K1} value. The process is iterative and may introduce artifacts at $\texttt{K1}=10$ and $\texttt{K1}=4.5$, where $\texttt{K1}=10$ is the default starting value, retained if no better fit is found, and $\texttt{K1}=4.5$ represents the lowest allowable value in the fitting routine\footnote{This information was provided by JPL SSD support staff at our request, after failing to find details in the literature or on the SSD website.}.  Comparing \texttt{K1} and $k_1$ highlights potential pitfalls in interpreting the JPL numbers at face value. Figure~\ref{fig:K1_vs_k1} shows that \texttt{K1} is generally underestimated relative to $k_1$. The Figure also reveals that $\texttt{K1}\geq 4.5$, the minimum permitted by the iterative fitting method used. Notably, the discrepancies between \texttt{K1} and $k_1$ appear to be independent of the comet's Oort group. 
    
    While a detailed investigation of causes of the differences between $k_1$ and \texttt{K1} is beyond the scope of this paper, some of the observed discrepancies may be explained by Figure~\ref{fig:incoming_vs_outgoing}, which shows that comets with larger $k_1$ values pre-perihelion tend to have significantly smaller $k_1$ pre-perihelion. Since \texttt{K1} is derived from data covering both orbital arcs, comets with steep brightening slopes (larger $k_1$) pre-perihelion are likely to have their \texttt{K1} values reduced by the influence of post-perihelion observations.

    \subsection{Implications for upcoming surveys and missions}

    This work was partially motivated by the upcoming LSST survey and Comet Interceptor (CI) mission. Developing CI without a predefined target presents clear challenges, making it essential to establish baseline expectations for potential targets. The target will likely be discovered by LSST, well beyond Jupiter \citep{2025Icar..42916443Inno}, and its brightening will be covered in great detail, albeit too late to influence most mission parameters.
    
    This paper scrutinizes long-standing assertions about LPC behavior using a large number of available measurements. Some of our findings are promising, particularly the evidence that new comets tend to exhibit consistent behavior. For instance, Figure~\ref{fig:delta_mag_vs_r} suggests that new comets discovered beyond Jupiter's orbit brighten in a relatively predictable manner compared to returning comets. Furthermore, if the typical rate of brightening of new comets continues to increase beyond the heliocentric distances probed by Figure~\ref{fig:kr_vs_r}, this may indicate that most new comets come from the most distant parts of the Oort cloud ($a>20\,000$ au) have not undergone significant processing and fading before reaching a near-Earth perihelion \citep{2022SciA....8M9130Kaib}.

    Table~\ref{tab:brightening_distributions} and Figure~\ref{fig:delta_mag_vs_r} can be used to convert the total magnitude of a comet discovered beyond Jupiter to an approximated, empirical probability for its total magnitude at 1 au. For instance, dynamically new comet \object{C/2021 T4} was tracked photometrically from $r=6.9$ au, inbound toward perihelion near $r=1.5$ au, as part of the LOOK survey \citep{2024PSJ.....5..273Holt}. Based on the initial heliocentric magnitude, $m_\text{helio}\approx 17.5$ mag, we would expect a median brightening of $\Delta m\approx 5.2$ mag (and 50\% chance of $4.8 < \Delta m < 6.0$ mag) upon reaching perihelion. According to the LOOK photometry, it reached $m_\text{helio}\approx 13$ mag, so it brightened by 4.5 magnitudes, only slightly below the statistical prediction. No other comet in the LOOK survey has observations from beyond 5 au to near 1 au. However, \object{C/2022 E3} (intermediate) is observed to brighten from approximately 14 mag at 4 au to 11.5 mag at 2 au (brightening of 2.5 mag) and to 10 mag at 1 au (brightening of 4 mag), while Table~\ref{tab:brightening_distributions} would predict median brightening of 2.55 and 5.8 mag in the same heliocentric ranges. Comet \object{C/2021 O3} (new) brightened 1.4 magnitudes (from 17.5 to 16.1 mag) between 4.2 and 3 au, compared to the a brightening of 1.62 mag between 4 and 3 au predicted by Table~\ref{tab:brightening_distributions}.

    \section{Conclusions}

    In this study, we investigated the secular brightening behavior of 272 long-period comets by analyzing a large sample of highly heterogeneous photometric observations. Using a robust selection process, we derived local and global brightening parameters, accounting for potential biases in the data. We examined how brightening rates vary with heliocentric distance and explored differences between dynamically new, intermediate, and old LPCs, and between pre- and post-perihelion observations. Below, we summarize the key conclusions of this work.

    \begin{enumerate}
        \item We find that dynamically new comets brighten more slowly than old comets. The difference in brightening rate is notable within 3 au from the Sun but becomes negligible beyond that distance.
        \item For new comets, the brightening rate varies with heliocentric distance becoming shallower as they approach the Sun: beyond 3 au they brighten at a rate around 12.8 mag per log au, whereas within 3 au from the Sun, their brightening rate decreases to below 7 mag per log au. Dynamically old comets brighten about 14 mag per log au in the entire range within 6 au of the Sun, displaying more scatter in their brightening behavior.
        \item Post-perihelion fading slopes are consistent across Oort groups, with little variation between new and returning comets.
        \item Dynamically new comets exhibit more uniform photometric behavior compared to returning comets, with a tighter correlation between brightening parameters $k_1$ and $m_1$. 
        \item Dynamically new comets are intrinsically brighter than old comets, possibly due to larger nuclei or higher activity levels.
    \end{enumerate}

    \section{Data availability}

    The full table of observations used in the paper, the table of processed strands and global curves and their properties, as well as the table of rejected strands and global curves are all available online \href{https://doi.org/10.5281/zenodo.15100031}{here}.

    \begin{acknowledgements}
      We thank David Jewitt for comments on the manuscript, and the referee for a detailed review which resulted in a paper with more robust statistics. PL acknowledges funding by Funda\c{c}\~{a}o para a Ci\^{e}ncia e a Tecnologia (FCT) through the research grants UIDB/04434/2020 (DOI: 10.54499/UIDB/04434/2020) and UIDP/04434/2020 (DOI: 10.54499/UIDP/04434/2020).  AGL was supported by CNES (mission Comet Interceptor) and funding from the European Research Council (ERC) under the European Union's Horizon 2020 research and innovation programme (Grant Agreement No 802699). RK would like to acknowledge the support from ``L'Oreal UNESCO For Women in Science'' National program for Bulgaria. RK acknowledges partial support by grant: K$\Pi$-06-H88/5 ``Physical properties and chemical composition of asteroids and comets - a key to increasing our knowledge of the Solar System origin and evolution.'' by the Bulgarian National Science Fund. LI acknowledges support by the Italian Space Agency (ASI) within the ASI-INAF agreements I/024/12/0 and 2020-4-HH.0.
    \end{acknowledgements}

\bibliographystyle{aa}
\bibliography{references}

\begin{thebibliography}{59}
\expandafter\ifx\csname natexlab\endcsname\relax\def\natexlab#1{#1}\fi

\bibitem[{{A'Hearn} {et~al.}(1995){A'Hearn}, {Millis}, {Schleicher}, {Osip}, \& {Birch}}]{1995Icar..118..223AHearn}
{A'Hearn}, M.~F., {Millis}, R.~C., {Schleicher}, D.~O., {Osip}, D.~J., \& {Birch}, P.~V. 1995, \icarus, 118, 223

\bibitem[{{Bauer} {et~al.}(2017){Bauer}, {Grav}, {Fern{\'a}ndez}, {Mainzer}, {Kramer}, {Masiero}, {Spahr}, {Nugent}, {Stevenson}, {Meech}, {Cutri}, {Lisse}, {Walker}, {Dailey}, {Rosser}, {Krings}, {Ruecker}, {Wright}, \& {NEOWISE Team}}]{2017AJ....154...53Bauer}
{Bauer}, J.~M., {Grav}, T., {Fern{\'a}ndez}, Y.~R., {et~al.} 2017, \aj, 154, 53

\bibitem[{{Betzler} {et~al.}(2023){Betzler}, {Diepvens}, \& {de Sousa}}]{2023MNRAS.526..246Betzler}
{Betzler}, A.~S., {Diepvens}, A., \& {de Sousa}, O.~F. 2023, \mnras, 526, 246

\bibitem[{{Bouziani} \& {Jewitt}(2022)}]{2022ApJ...924...37Bouziani}
{Bouziani}, N. \& {Jewitt}, D. 2022, \apj, 924, 37

\bibitem[{{Dello Russo} {et~al.}(2016){Dello Russo}, {Kawakita}, {Vervack}, \& {Weaver}}]{2016Icar..278..301DelloRusso}
{Dello Russo}, N., {Kawakita}, H., {Vervack}, R.~J., \& {Weaver}, H.~A. 2016, \icarus, 278, 301

\bibitem[{{Delsemme}(1982)}]{1982come.coll...85Delsemme}
{Delsemme}, A.~H. 1982, in IAU Colloq. 61: Comet Discoveries, Statistics, and Observational Selection, ed. L.~L. {Wilkening}, 85--130

\bibitem[{{Duncan} {et~al.}(1987){Duncan}, {Quinn}, \& {Tremaine}}]{1987AJ.....94.1330Duncan}
{Duncan}, M., {Quinn}, T., \& {Tremaine}, S. 1987, \aj, 94, 1330

\bibitem[{{Fern{\'a}ndez} {et~al.}(1999){Fern{\'a}ndez}, {Tancredi}, {Rickman}, \& {Licandro}}]{1999A&A...352..327Fernandez}
{Fern{\'a}ndez}, J.~A., {Tancredi}, G., {Rickman}, H., \& {Licandro}, J. 1999, \aap, 352, 327

\bibitem[{{Francis}(2005)}]{2005ApJ...635.1348Francis}
{Francis}, P.~J. 2005, \apj, 635, 1348

\bibitem[{{Fulle} {et~al.}(2020){Fulle}, {Blum}, \& {Rotundi}}]{2020A&A...636L...3Fulle}
{Fulle}, M., {Blum}, J., \& {Rotundi}, A. 2020, \aap, 636, L3

\bibitem[{{Ginsburg} {et~al.}(2019){Ginsburg}, {Sip{\H o}cz}, {Brasseur}, {Cowperthwaite}, {Craig}, {Deil}, {Guillochon}, {Guzman}, {Liedtke}, {Lian Lim}, {Lockhart}, {Mommert}, {Morris}, {Norman}, {Parikh}, {Persson}, {Robitaille}, {Segovia}, {Singer}, {Tollerud}, {de Val-Borro}, {Valtchanov}, {Woillez}, {The Astroquery collaboration}, \& {a subset of the astropy collaboration}}]{2019AJ....157...98Ginsburg}
{Ginsburg}, A., {Sip{\H o}cz}, B.~M., {Brasseur}, C.~E., {et~al.} 2019, \aj, 157, 98

\bibitem[{{Gkotsinas} {et~al.}(2024){Gkotsinas}, {Nesvorn{\'y}}, {Guilbert-Lepoutre}, {Raymond}, \& {Kaib}}]{2024PSJ.....5..243Gkotsinas}
{Gkotsinas}, A., {Nesvorn{\'y}}, D., {Guilbert-Lepoutre}, A., {Raymond}, S.~N., \& {Kaib}, N. 2024, Planetary Science Journal, 5, 243

\bibitem[{{Green}(1995)}]{1995ICQ....17..168Green}
{Green}, D. W.~E. 1995, International Comet Quarterly, 17, 168

\bibitem[{{Harrington Pinto} {et~al.}(2022){Harrington Pinto}, {Womack}, {Fernandez}, \& {Bauer}}]{2022PSJ.....3..247HPinto}
{Harrington Pinto}, O., {Womack}, M., {Fernandez}, Y., \& {Bauer}, J. 2022, Planetary Science Journal, 3, 247

\bibitem[{{Heisler} \& {Tremaine}(1986)}]{1986Icar...65...13Heisler}
{Heisler}, J. \& {Tremaine}, S. 1986, \icarus, 65, 13

\bibitem[{{Heisler} {et~al.}(1987){Heisler}, {Tremaine}, \& {Alcock}}]{1987Icar...70..269Heisler}
{Heisler}, J., {Tremaine}, S., \& {Alcock}, C. 1987, \icarus, 70, 269

\bibitem[{{Higuchi} \& {Kokubo}(2015)}]{2015AJ....150...26Higuchi}
{Higuchi}, A. \& {Kokubo}, E. 2015, \aj, 150, 26

\bibitem[{{Higuchi} {et~al.}(2007){Higuchi}, {Kokubo}, {Kinoshita}, \& {Mukai}}]{2007AJ....134.1693Higuchi}
{Higuchi}, A., {Kokubo}, E., {Kinoshita}, H., \& {Mukai}, T. 2007, \aj, 134, 1693

\bibitem[{{Hodapp} {et~al.}(2004){Hodapp}, {Kaiser}, {Aussel}, {Burgett}, {Chambers}, {Chun}, {Dombeck}, {Douglas}, {Hafner}, {Heasley}, {Hoblitt}, {Hude}, {Isani}, {Jedicke}, {Jewitt}, {Laux}, {Luppino}, {Lupton}, {Maberry}, {Magnier}, {Mannery}, {Monet}, {Morgan}, {Onaka}, {Price}, {Ryan}, {Siegmund}, {Szapudi}, {Tonry}, {Wainscoat}, \& {Waterson}}]{2004AN....325..636Hodapp}
{Hodapp}, K.~W., {Kaiser}, N., {Aussel}, H., {et~al.} 2004, Astronomische Nachrichten, 325, 636

\bibitem[{{Holt} {et~al.}(2024){Holt}, {Knight}, {Kelley}, {Lister}, {Ye}, {Snodgrass}, {Opitom}, {Kokotanekova}, {Schwamb}, {Dobson}, {Bannister}, {Micheli}, {Milam}, {Richardson}, {Gomez}, {Chatelain}, \& {Greenstreet}}]{2024PSJ.....5..273Holt}
{Holt}, C.~E., {Knight}, M.~M., {Kelley}, M. S.~P., {et~al.} 2024, Planetary Science Journal, 5, 273

\bibitem[{Huber {et~al.}(1981)Huber, Wiley, \& InterScience}]{1981Robuststatistics.Huber}
Huber, P., Wiley, J., \& InterScience, W. 1981, {Robust statistics} (Wiley New York)

\bibitem[{{Hughes}(1988)}]{1988MNRAS.234..173Hughes}
{Hughes}, D.~W. 1988, \mnras, 234, 173

\bibitem[{{Hughes} {et~al.}(1993){Hughes}, {McBride}, {Boswell}, \& {Jalowiczor}}]{1993MNRAS.263..247Hughes}
{Hughes}, D.~W., {McBride}, N., {Boswell}, J., \& {Jalowiczor}, P. 1993, \mnras, 263, 247

\bibitem[{{Inno} {et~al.}(2025){Inno}, {Scuderi}, {Bertini}, {Fulle}, {Mazzotta Epifani}, {Della Corte}, {Piccirillo}, {Vanzanella}, {Lacerda}, {Grappasonni}, {Ammanito}, {Sindoni}, \& {Rotundi}}]{2025Icar..42916443Inno}
{Inno}, L., {Scuderi}, M., {Bertini}, I., {et~al.} 2025, \icarus, 429, 116443

\bibitem[{{Ivezi{\'c}} {et~al.}(2019){Ivezi{\'c}}, {Kahn}, {Tyson}, {Abel}, {Acosta}, {Allsman}, {Alonso}, {AlSayyad}, {Anderson}, {Andrew}, {Angel}, {Angeli}, {Ansari}, {Antilogus}, {Araujo}, {Armstrong}, {Arndt}, {Astier}, {Aubourg}, {Auza}, {Axelrod}, {Bard}, {Barr}, {Barrau}, {Bartlett}, {Bauer}, {Bauman}, {Baumont}, {Bechtol}, {Bechtol}, {Becker}, {Becla}, {Beldica}, {Bellavia}, {Bianco}, {Biswas}, {Blanc}, {Blazek}, {Blandford}, {Bloom}, {Bogart}, {Bond}, {Booth}, {Borgland}, {Borne}, {Bosch}, {Boutigny}, {Brackett}, {Bradshaw}, {Brandt}, {Brown}, {Bullock}, {Burchat}, {Burke}, {Cagnoli}, {Calabrese}, {Callahan}, {Callen}, {Carlin}, {Carlson}, {Chandrasekharan}, {Charles-Emerson}, {Chesley}, {Cheu}, {Chiang}, {Chiang}, {Chirino}, {Chow}, {Ciardi}, {Claver}, {Cohen-Tanugi}, {Cockrum}, {Coles}, {Connolly}, {Cook}, {Cooray}, {Covey}, {Cribbs}, {Cui}, {Cutri}, {Daly}, {Daniel}, {Daruich}, {Daubard}, {Daues}, {Dawson}, {Delgado}, {Dellapenna}, {de Peyster}, {de Val-Borro}, {Digel}, {Doherty}, {Dubois},
  {Dubois-Felsmann}, {Durech}, {Economou}, {Eifler}, {Eracleous}, {Emmons}, {Fausti Neto}, {Ferguson}, {Figueroa}, {Fisher-Levine}, {Focke}, {Foss}, {Frank}, {Freemon}, {Gangler}, {Gawiser}, {Geary}, {Gee}, {Geha}, {Gessner}, {Gibson}, {Gilmore}, {Glanzman}, {Glick}, {Goldina}, {Goldstein}, {Goodenow}, {Graham}, {Gressler}, {Gris}, {Guy}, {Guyonnet}, {Haller}, {Harris}, {Hascall}, {Haupt}, {Hernandez}, {Herrmann}, {Hileman}, {Hoblitt}, {Hodgson}, {Hogan}, {Howard}, {Huang}, {Huffer}, {Ingraham}, {Innes}, {Jacoby}, {Jain}, {Jammes}, {Jee}, {Jenness}, {Jernigan}, {Jevremovi{\'c}}, {Johns}, {Johnson}, {Johnson}, {Jones}, {Juramy-Gilles}, {Juri{\'c}}, {Kalirai}, {Kallivayalil}, {Kalmbach}, {Kantor}, {Karst}, {Kasliwal}, {Kelly}, {Kessler}, {Kinnison}, {Kirkby}, {Knox}, {Kotov}, {Krabbendam}, {Krughoff}, {Kub{\'a}nek}, {Kuczewski}, {Kulkarni}, {Ku}, {Kurita}, {Lage}, {Lambert}, {Lange}, {Langton}, {Le Guillou}, {Levine}, {Liang}, {Lim}, {Lintott}, {Long}, {Lopez}, {Lotz}, {Lupton}, {Lust}, {MacArthur}, {Mahabal},
  {Mandelbaum}, {Markiewicz}, {Marsh}, {Marshall}, {Marshall}, {May}, {McKercher}, {McQueen}, {Meyers}, {Migliore}, {Miller}, {Mills}, {Miraval}, {Moeyens}, {Moolekamp}, {Monet}, {Moniez}, {Monkewitz}, {Montgomery}, {Morrison}, {Mueller}, {Muller}, {Mu{\~n}oz Arancibia}, {Neill}, {Newbry}, {Nief}, {Nomerotski}, {Nordby}, {O'Connor}, {Oliver}, {Olivier}, {Olsen}, {O'Mullane}, {Ortiz}, {Osier}, {Owen}, {Pain}, {Palecek}, {Parejko}, {Parsons}, {Pease}, {Peterson}, {Peterson}, {Petravick}, {Libby Petrick}, {Petry}, {Pierfederici}, {Pietrowicz}, {Pike}, {Pinto}, {Plante}, {Plate}, {Plutchak}, {Price}, {Prouza}, {Radeka}, {Rajagopal}, {Rasmussen}, {Regnault}, {Reil}, {Reiss}, {Reuter}, {Ridgway}, {Riot}, {Ritz}, {Robinson}, {Roby}, {Roodman}, {Rosing}, {Roucelle}, {Rumore}, {Russo}, {Saha}, {Sassolas}, {Schalk}, {Schellart}, {Schindler}, {Schmidt}, {Schneider}, {Schneider}, {Schoening}, {Schumacher}, {Schwamb}, {Sebag}, {Selvy}, {Sembroski}, {Seppala}, {Serio}, {Serrano}, {Shaw}, {Shipsey}, {Sick}, {Silvestri},
  {Slater}, {Smith}, {Smith}, {Sobhani}, {Soldahl}, {Storrie-Lombardi}, {Stover}, {Strauss}, {Street}, {Stubbs}, {Sullivan}, {Sweeney}, {Swinbank}, {Szalay}, {Takacs}, {Tether}, {Thaler}, {Thayer}, {Thomas}, {Thornton}, {Thukral}, {Tice}, {Trilling}, {Turri}, {Van Berg}, {Vanden Berk}, {Vetter}, {Virieux}, {Vucina}, {Wahl}, {Walkowicz}, {Walsh}, {Walter}, {Wang}, {Wang}, {Warner}, {Wiecha}, {Willman}, {Winters}, {Wittman}, {Wolff}, {Wood-Vasey}, {Wu}, {Xin}, {Yoachim}, \& {Zhan}}]{2019ApJ...873..111Ivezic}
{Ivezi{\'c}}, {\v{Z}}., {Kahn}, S.~M., {Tyson}, J.~A., {et~al.} 2019, \apj, 873, 111

\bibitem[{{Jewitt}(1991)}]{1991ASSL..167...19Jewitt}
{Jewitt}, D. 1991, in Astrophysics and Space Science Library, Vol. 167, IAU Colloq. 116: Comets in the post-Halley era, ed. J.~{Newburn}, R.~L., M.~{Neugebauer}, \& J.~{Rahe}, 19

\bibitem[{{Jewitt}(2022)}]{2022AJ....164..158Jewitt}
{Jewitt}, D. 2022, \aj, 164, 158

\bibitem[{{Jewitt} {et~al.}(2019){Jewitt}, {Agarwal}, {Hui}, {Li}, {Mutchler}, \& {Weaver}}]{2019AJ....157...65Jewitt}
{Jewitt}, D., {Agarwal}, J., {Hui}, M.-T., {et~al.} 2019, \aj, 157, 65

\bibitem[{{Jewitt} {et~al.}(2021){Jewitt}, {Kim}, {Mutchler}, {Agarwal}, {Li}, \& {Weaver}}]{2021AJ....161..188Jewitt}
{Jewitt}, D., {Kim}, Y., {Mutchler}, M., {et~al.} 2021, \aj, 161, 188

\bibitem[{{Jones} {et~al.}(2024){Jones}, {Snodgrass}, {Tubiana}, {K{\"u}ppers}, {Kawakita}, {Lara}, {Agarwal}, {Andr{\'e}}, {Attree}, {Auster}, {Bagnulo}, {Bannister}, {Beth}, {Bowles}, {Coates}, {Colangeli}, {Corral van Damme}, {Da Deppo}, {De Keyser}, {Della Corte}, {Edberg}, {El-Maarry}, {Faggi}, {Fulle}, {Funase}, {Galand}, {Goetz}, {Groussin}, {Guilbert-Lepoutre}, {Henri}, {Kasahara}, {Kereszturi}, {Kidger}, {Knight}, {Kokotanekova}, {Kolmasova}, {Kossacki}, {K{\"u}hrt}, {Kwon}, {La Forgia}, {Levasseur-Regourd}, {Lippi}, {Longobardo}, {Marschall}, {Morawski}, {Mu{\~n}oz}, {N{\"a}sil{\"a}}, {Nilsson}, {Opitom}, {Pajusalu}, {Pommerol}, {Prech}, {Rando}, {Ratti}, {Rothkaehl}, {Rotundi}, {Rubin}, {Sakatani}, {S{\'a}nchez}, {Simon Wedlund}, {Stankov}, {Thomas}, {Toth}, {Villanueva}, {Vincent}, {Volwerk}, {Wurz}, {Wielders}, {Yoshioka}, {Aleksiejuk}, {Alvarez}, {Amoros}, {Aslam}, {Atamaniuk}, {Baran}, {Barci{\'n}ski}, {Beck}, {Behnke}, {Berglund}, {Bertini}, {Bieda}, {Binczyk}, {Busch}, {Cacovean}, {Capria},
  {Carr}, {Castro Mar{\'\i}n}, {Ceriotti}, {Chioetto}, {Chuchra-Konrad}, {Cocola}, {Colin}, {Crews}, {Cripps}, {Cupido}, {Dassatti}, {Davidsson}, {De Roche}, {Deca}, {Del Togno}, {Dhooghe}, {Donaldson Hanna}, {Eriksson}, {Fedorov}, {Fern{\'a}ndez-Valenzuela}, {Ferretti}, {Floriot}, {Frassetto}, {Fredriksson}, {Garnier}, {Gawe{\l}}, {G{\'e}not}, {Gerber}, {Glassmeier}, {Granvik}, {Grison}, {Gunell}, {Hachemi}, {Hagen}, {Hajra}, {Harada}, {Hasiba}, {Haslebacher}, {Herranz De La Revilla}, {Hestroffer}, {Hewagama}, {Holt}, {Hviid}, {Iakubivskyi}, {Inno}, {Irwin}, {Ivanovski}, {Jansky}, {Jernej}, {Jeszenszky}, {Jimen{\'e}z}, {Jorda}, {Kama}, {Kameda}, {Kelley}, {Klepacki}, {Kohout}, {Kojima}, {Kowalski}, {Kuwabara}, {Ladno}, {Laky}, {Lammer}, {Lan}, {Lavraud}, {Lazzarin}, {Le Duff}, {Lee}, {Lesniak}, {Lewis}, {Lin}, {Lister}, {Lowry}, {Magnes}, {Markkanen}, {Martinez Navajas}, {Martins}, {Matsuoka}, {Matyjasiak}, {Mazelle}, {Mazzotta Epifani}, {Meier}, {Michaelis}, {Micheli}, {Migliorini}, {Millet}, {Moreno},
  {Mottola}, {Moutounaick}, {Muinonen}, {M{\"u}ller}, {Murakami}, {Murata}, {Myszka}, {Nakajima}, {Nemeth}, {Nikolajev}, {Nordera}, {Ohlsson}, {Olesk}, {Ottacher}, {Ozaki}, {Oziol}, {Patel}, {Savio Paul}, {Penttil{\"a}}, {Pernechele}, {Peterson}, {Petraglio}, {Piccirillo}, {Plaschke}, {Polak}, {Postberg}, {Proosa}, {Protopapa}, {Puccio}, {Ranvier}, {Raymond}, {Richter}, {Rieder}, {Rigamonti}, {Ruiz Rodriguez}, {Santolik}, {Sasaki}, {Schr{\"o}dter}, {Shirley}, {Slavinskis}, {Sodor}, {Soucek}, {Stephenson}, {St{\"o}ckli}, {Szewczyk}, {Troznai}, {Uhlir}, {Usami}, {Valavanoglou}, {Vaverka}, {Wang}, {Wang}, {Wattieaux}, {Wieser}, {Wolf}, {Yano}, {Yoshikawa}, {Zakharov}, {Zawistowski}, {Zuppella}, {Rinaldi}, \& {Ji}}]{2024SSRv..220....9Jones}
{Jones}, G.~H., {Snodgrass}, C., {Tubiana}, C., {et~al.} 2024, \ssr, 220, 9

\bibitem[{{Kaib}(2022)}]{2022SciA....8M9130Kaib}
{Kaib}, N.~A. 2022, Science Advances, 8, eabm9130

\bibitem[{{Kaib} \& {Quinn}(2009)}]{2009Sci...325.1234Kaib}
{Kaib}, N.~A. \& {Quinn}, T. 2009, Science, 325, 1234

\bibitem[{{Keller} \& {K{\"u}hrt}(2020)}]{2020SSRv..216...14Keller}
{Keller}, H.~U. \& {K{\"u}hrt}, E. 2020, \ssr, 216, 14

\bibitem[{{Kr{\'o}likowska} \& {Dybczy{\'n}ski}(2020)}]{2020A&A...640A..97Krolikowska}
{Kr{\'o}likowska}, M. \& {Dybczy{\'n}ski}, P.~A. 2020, \aap, 640, A97

\bibitem[{{Larson} {et~al.}(2003){Larson}, {Beshore}, {Hill}, {Christensen}, {McLean}, {Kolar}, {McNaught}, \& {Garradd}}]{2003DPS....35.3604Larson}
{Larson}, S., {Beshore}, E., {Hill}, R., {et~al.} 2003, in AAS/Division for Planetary Sciences Meeting Abstracts, Vol.~35, AAS/Division for Planetary Sciences Meeting Abstracts \#35, 36.04

\bibitem[{{Maggiolo} {et~al.}(2020){Maggiolo}, {Gronoff}, {Cessateur}, {Moore}, {Airapetian}, {De Keyser}, {Dhooghe}, {Gibbons}, {Gunell}, {Mertens}, {Rubin}, \& {Hosseini}}]{2020ApJ...901..136Maggiolo}
{Maggiolo}, R., {Gronoff}, G., {Cessateur}, G., {et~al.} 2020, \apj, 901, 136

\bibitem[{{Mainzer} {et~al.}(2011){Mainzer}, {Bauer}, {Grav}, {Masiero}, {Cutri}, {Dailey}, {Eisenhardt}, {McMillan}, {Wright}, {Walker}, {Jedicke}, {Spahr}, {Tholen}, {Alles}, {Beck}, {Brandenburg}, {Conrow}, {Evans}, {Fowler}, {Jarrett}, {Marsh}, {Masci}, {McCallon}, {Wheelock}, {Wittman}, {Wyatt}, {DeBaun}, {Elliott}, {Elsbury}, {Gautier}, {Gomillion}, {Leisawitz}, {Maleszewski}, {Micheli}, \& {Wilkins}}]{2011ApJ...731...53Mainzer}
{Mainzer}, A., {Bauer}, J., {Grav}, T., {et~al.} 2011, \apj, 731, 53

\bibitem[{{Malamud} {et~al.}(2022){Malamud}, {Landeck}, {Bischoff}, {Kreuzig}, {Perets}, {Gundlach}, \& {Blum}}]{2022MNRAS.514.3366Malamud}
{Malamud}, U., {Landeck}, W.~A., {Bischoff}, D., {et~al.} 2022, \mnras, 514, 3366

\bibitem[{{Marshall} {et~al.}(2019){Marshall}, {Rezac}, {Hartogh}, {Zhao}, \& {Attree}}]{2019A&A...623A.120Marschall}
{Marshall}, D., {Rezac}, L., {Hartogh}, P., {Zhao}, Y., \& {Attree}, N. 2019, \aap, 623, A120

\bibitem[{{Mazzotta Epifani} {et~al.}(2014){Mazzotta Epifani}, {Perna}, {Di Fabrizio}, {Dall'Ora}, {Palumbo}, {Snodgrass}, {Licandro}, {Della Corte}, \& {Tozzi}}]{2014A&A...561A...6MazzottaEpifani}
{Mazzotta Epifani}, E., {Perna}, D., {Di Fabrizio}, L., {et~al.} 2014, \aap, 561, A6

\bibitem[{{Meech} \& {Svoren}(2004)}]{2004come.book..317Meech}
{Meech}, K.~J. \& {Svoren}, J. 2004, in Comets II, ed. M.~C. {Festou}, H.~U. {Keller}, \& H.~A. {Weaver} (University of Arizona Press), 317

\bibitem[{{Meisel}(1970)}]{1970AJ.....75..252Meisel}
{Meisel}, D.~D. 1970, \aj, 75, 252

\bibitem[{{Meisel} \& {Morris}(1976)}]{1976NASSP.393..410Meisel}
{Meisel}, D.~D. \& {Morris}, C.~S. 1976, in NASA Special Publication, ed. B.~{Donn}, M.~{Mumma}, W.~{Jackson}, M.~{A'Hearn}, \& R.~S. {Harrington}, Vol. 393 (NASA), 410--444

\bibitem[{{Meisel} \& {Morris}(1982)}]{1982come.coll..413Meisel}
{Meisel}, D.~M. \& {Morris}, C.~S. 1982, in IAU Colloq. 61: Comet Discoveries, Statistics, and Observational Selection, ed. L.~L. {Wilkening}, 413--432

\bibitem[{{Oort}(1950)}]{1950BAN....11...91Oort}
{Oort}, J.~H. 1950, \bain, 11, 91

\bibitem[{{Oort} \& {Schmidt}(1951)}]{1951BAN....11..259Oort}
{Oort}, J.~H. \& {Schmidt}, M. 1951, \bain, 11, 259

\bibitem[{{Pfalzner} {et~al.}(2024){Pfalzner}, {Govind}, \& {Portegies Zwart}}]{2024NatAs...8.1380Pfalzner}
{Pfalzner}, S., {Govind}, A., \& {Portegies Zwart}, S. 2024, Nature Astronomy, 8, 1380

\bibitem[{{Robinson} {et~al.}(2024){Robinson}, {Malamud}, {Opitom}, {Perets}, \& {Blum}}]{2024MNRAS.531..859Robinson}
{Robinson}, J.~E., {Malamud}, U., {Opitom}, C., {Perets}, H., \& {Blum}, J. 2024, \mnras, 531, 859

\bibitem[{{S{\'a}rneczky} {et~al.}(2016){S{\'a}rneczky}, {Szab{\'o}}, {Cs{\'a}k}, {Kelemen}, {Marschalk{\'o}}, {P{\'a}l}, {Szak{\'a}ts}, {Szalai}, {Szegedi-Elek}, {Sz{\'e}kely}, {Vida}, {Vink{\'o}}, \& {Kiss}}]{2016AJ....152..220Sarneczky}
{S{\'a}rneczky}, K., {Szab{\'o}}, G.~M., {Cs{\'a}k}, B., {et~al.} 2016, \aj, 152, 220

\bibitem[{{Smith} {et~al.}(2020){Smith}, {Smartt}, {Young}, {Tonry}, {Denneau}, {Flewelling}, {Heinze}, {Weiland}, {Stalder}, {Rest}, {Stubbs}, {Anderson}, {Chen}, {Clark}, {Do}, {F{\"o}rster}, {Fulton}, {Gillanders}, {McBrien}, {O'Neill}, {Srivastav}, \& {Wright}}]{2020PASP..132h5002Smith}
{Smith}, K.~W., {Smartt}, S.~J., {Young}, D.~R., {et~al.} 2020, \pasp, 132, 085002

\bibitem[{{Snodgrass} \& {Holt}(2024)}]{2024EPSC...17..324Snodgrass}
{Snodgrass}, C. \& {Holt}, C. 2024, in European Planetary Science Congress, EPSC2024--324

\bibitem[{{Sosa} \& {Fern{\'a}ndez}(2011)}]{2011MNRAS.416..767Sosa}
{Sosa}, A. \& {Fern{\'a}ndez}, J.~A. 2011, \mnras, 416, 767

\bibitem[{{Stokes} {et~al.}(2000){Stokes}, {Evans}, {Viggh}, {Shelly}, \& {Pearce}}]{2000Icar..148...21Stokes}
{Stokes}, G.~H., {Evans}, J.~B., {Viggh}, H. E.~M., {Shelly}, F.~C., \& {Pearce}, E.~C. 2000, \icarus, 148, 21

\bibitem[{{van Woerkom}(1948)}]{1948BAN....10..445VanWoerkom}
{van Woerkom}, A.~J.~J. 1948, \bain, 10, 445

\bibitem[{{Weissman}(1980)}]{1980A&A....85..191Weissman}
{Weissman}, P.~R. 1980, \aap, 85, 191

\bibitem[{{Whipple}(1978)}]{1978M&P....18..343Whipple}
{Whipple}, F.~L. 1978, Moon and Planets, 18, 343

\bibitem[{{Whipple}(1992)}]{1992acm..proc..633Whipple}
{Whipple}, F.~L. 1992, in Asteroids, Comets, Meteors 1991, ed. A.~W. {Harris} \& E.~{Bowell}, 633

\bibitem[{{Wiegert} \& {Tremaine}(1999)}]{1999Icar..137...84Wiegert}
{Wiegert}, P. \& {Tremaine}, S. 1999, \icarus, 137, 84

\bibitem[{{Yang} {et~al.}(2021){Yang}, {Jewitt}, {Zhao}, {Jiang}, {Ye}, \& {Chen}}]{2021ApJ...914L..17Yang}
{Yang}, B., {Jewitt}, D., {Zhao}, Y., {et~al.} 2021, \apjl, 914, L17

\end{thebibliography}

\begin{appendix}
    \section{Sample of observations}
    \label{app:curves}

\begin{figure*}
        \centering
        \includegraphics[width=17cm]{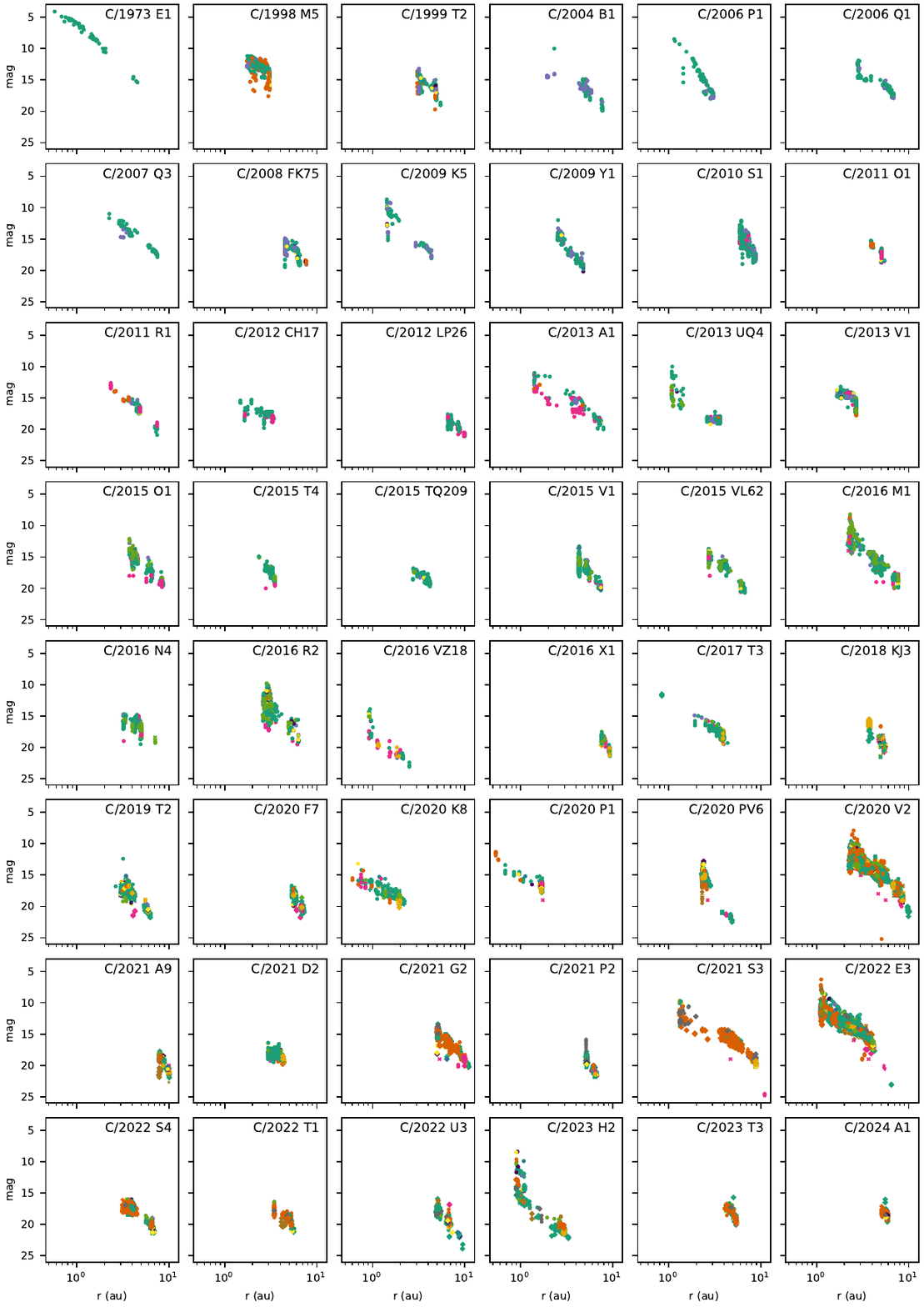}
            \caption{Larger subset of observations grouped by comet. Colors correspond to different observatories and symbols to different photometric bands.}
        \label{fig:large_sample_curves}
    \end{figure*}

    Figure~\ref{fig:large_sample_curves} shows a larger sample of curves. For the full dataset of observations follow this \href{https://doi.org/10.5281/zenodo.15100031}{online link}.
    
\end{appendix}

\end{document}